\documentclass[10pt,aps,prd,twocolumn,superscriptaddress,%
preprintnumbers,showpacs,amsmath,nofootinbib,a4paper]{revtex4}
\usepackage{graphicx}
\usepackage{amsmath}
\renewcommand{\ol}{\overline}
\newcommand{\Tr}{\mbox{Tr}}
\newcommand{\eqn}[1]{&\hspace{-0.4em}#1\hspace{-0.4em}&}
\newcommand{\vev}[1]{\mbox{$\langle #1 \rangle$}}
\newcommand{\abs}[1]{\mbox{$\left| #1 \right|$}}
\newcommand{\ie}{\hbox{\it i.e.}{}}
\newcommand{\eg}{\hbox{\it e.g.}{}}
\newcommand{\etc}{\hbox{\it etc.}{}}

\begin{document}
\preprint{IC/2004/84}
\preprint{hep-ph/0409147}
\title{Masses and Mixing of Quarks and Leptons
in Product-Group Unification}
\author{Takehiko Asaka}
\affiliation{Institute of Theoretical Physics, 
Swiss Federal Institute of Technology, 
CH-1015 Lausanne, Switzerland}
\author{Y.~Takanishi}
\affiliation{The Abdus Salam International Centre for Theoretical Physics,
Strada Costiera 11, I-34100 Trieste, Italy}
\date{September 13, 2004}
%
%%%%%%%%%%%%%%%%%%%%%%%%%%%%%%%%%%%%%%%%%%%%%%%%%%%%%%%%%%%%%%%%%%%%%
\begin{abstract}
  We discuss a supersymmetric unified model based on a product gauge
  group SU(5)$\times$SU(5)$\times$SU(5), where the gauge symmetry
  breaking is achieved without the adjoint or higher-dimensional Higgs
  field, and the doublet-triplet splitting in the Higgs masses is
  realized by the use of the discrete symmetry.  In this article we
  present an explicit model for realistic fermion masses with the
  discrete symmetries $Z_7\times Z_2$.  It is shown that all the
  observed masses and mixing angles for quarks and leptons, including
  neutrinos, are well described by the breaking of the symmetries
  imposed in the model.  Especially, the maximal and large mixing
  angles in the atmospheric and solar neutrino oscillations are
  obtained as the most preferred values, and the typical value of the
  neutrino mixing element $U_{e 3}$ is 0.1--0.3. We also point out the
  non-trivial relations among the $\mu$-parameter for the Higgs mass,
  the charged fermion hierarchies, and the neutrino masses.  These
  relations suggest that the scale of $\mu$ is of order of the weak
  scale.
\end{abstract}
%%%%%%%%%%%%%%%%%%%%%%%%%%%%%%%%%%%%%%%%%%%%%%%%%%%%%%%%%%%%%%%%%%%%%
% 
%\keywords{}
%
\maketitle
%%%%%%%%%%%%%%%%%%%%%%%%%%%%%%%%%%%%%%%%%%%%%%%%%%%%%%%%%%%%%%%%%%%%%
\section{Introduction}
\indent
The unification of strong, weak and electromagnetic interactions is a
very attractive idea~\cite{Georgi:1974sy}.  The gauge coupling
unification supports strongly this grand unified theories (GUTs) with
a supersymmetry at low energy.  The simplest model is based on an
SU(5) group~\cite{SUSYGUT}.  In spite of the theoretical appeal, there
have been several puzzles in construction of supersymmetric GUT
models: The first puzzle is the Higgs sector.  In general, the very
complicated and innovative Higgs sector should be considered.  The GUT
gauge breaking usually requires the Higgs field(s) in the adjoint or
higher-dimensional representation(s).  In particular, the most
annoying difficulty is the mass splitting between Higgs weak-doublets
and color-triplets.  Various mechanisms for this problem have been
proposed~\cite{MissingPartner,MissingVEV,SlidingSinglet,GIFT,Orbifold}.

The second one concerns the matter sector.  Quarks and leptons are
unified into irreducible representation(s) of the GUT group.  This
matter unification ensures the electric charge quantization, but also
predicts a promising mass relation, $m_b=m_\tau$, in the simplest
SU(5) model.  However, it predicts also the unwanted mass
relations, $m_s=m_\mu$ and $m_d=m_e$, which are badly broken in
nature.  Moreover, recent neutrino
experiments~\cite{ATM,SOL,SNO,KamLAND} reveal that the neutrinos own
non-zero masses which are much smaller than the electron mass, and
remarkably that the mixing angles in the atmospheric and solar
neutrino oscillations are both large.  The rest angle is
only limited by experiment~\cite{CHOOZ}.  We then have to
explain small and large mixing angles in the quark-lepton sector
at the same time.  The observed fermion properties
may also require the GUT models to be complicated.

One way to evade these issues is to give up the fundamental assumption
-- a simple group --, $\ie$, to go into the product group like
$G\times G $ or $G\times G\times G$, where $G$ denotes the GUT gauge
groups~\cite{Barbieri:1994jq,Barbieri:1994cx,Frampton:1995gu,Mohapatra:1996fu,Barr:1996kp,Maslikov:1996gn,Chou:1998pr,Witten:2001bf,Dine:2002se}.
As pointed out in Ref.~\cite{Barbieri:1994jq} the class of these
models, called the product-group unification, still preserves the nice
features of the GUTs.  Namely, the standard gauge group is contained
in the diagonal subgroup, and hence the coupling unification is
realized at tree-level.  In addition, quarks and leptons are unified
in each GUT group $G$, and the electric charge quantization is also
explained.

There are three major advantages in the product-group unification:
First, the breaking of gauge symmetry is accomplished without
introducing the Higgs fields in the adjoint or higher-dimensional
representations~\cite{Barbieri:1994jq}.  This seems very promising for
the connection to the string theory.  In fact, it is known that there
is no state in the adjoint or higher representation in the string
spectrum with the affine level one~\cite{Dienes:1996du}. Second, the
doublet-triplet problem may receive a natural solution by the use of
the discrete
symmetry~\cite{Barbieri:1994jq,Barr:1996kp,Chou:1998pr,Witten:2001bf,Dine:2002se}.\footnote{In
  the different class of models with a product group the unbroken
  $R$-symmetry is used for the doublet-triplet
  splitting~\cite{Yanagida:1994vq}.}  Finally, we may gain important
insights to the flavor physics.  It is beyond the scope of the usual
GUTs (as well as the standard model) to understand the origin of the
hierarchical structure in the Yukawa couplings, which are essentially
free parameters.  On the other hand, the mass hierarchy between
different flavors may be explained by the breaking of the product
group when they belong to different GUT group
factors~\cite{Barbieri:1994cx}.  This illustrates the idea of
Refs.~\cite{Froggatt:tt,Nielsen:2002cw} in the unified theories.
Moreover, the discrete symmetry for the doublet-triplet splitting can
be utilized as the horizontal flavor symmetry, which is also crucial
for the suppression of the dimension-five operators to avoid a rapid
proton
decay~\cite{Barbieri:1994cx,Chou:1998pr,Witten:2001bf,Dine:2002se}.

In this article, we present an explicit model for realistic fermion
masses based on the product group SU(5)$\times$SU(5)$\times$SU(5).
Especially, we emphasize on the neutrino properties.  Although the
similar analyses have been done in
Refs.~\cite{Barbieri:1994cx,Chou:1998pr}, the model in
Ref.~\cite{Barbieri:1994cx} leads to the small angles for neutrino
mixing, and in Ref.~\cite{Chou:1998pr} the neutrino physics has not
been discussed.  Our model is constructed by imposing the discrete
symmetries $Z_7\times Z_2$, where the $Z_7$ symmetry is introduced to
solve the doublet-triplet splitting.  The observed masses and flavor
mixing of quarks and leptons, including neutrinos, are well described
by the breaking of the product gauge group together with these discrete
symmetries.  It is shown that the $Z_7$ breaking is crucial for generating
({\it i}) an effective $\mu$-term for the Higgs doublets, 
({\it ii}) the charged fermion mass hierarchies,
and ({\it iii}) the Majorana masses for right-handed neutrinos.%
%%%%%%%%%%%%%%%%%%%%%%%%%%%%%
\footnote{The relation between the large Majorana masses 
for right-handed neutrinos and the breaking of the horizontal 
global symmetry has been discussed in Ref.~\cite{Yanagida:1979gs}.}
%%%%%%%%%%%%%%%%%%%%%%%%%%%%%
We point out that the $\mu$-parameter at the weak scale
is suggested from not only the observed masses of charged fermions,
but also the neutrino masses indicated by the oscillation experiments.

This article is organized as follows: in the next section, we first
present our model based on a product group
SU(5)$\times$SU(5)$\times$SU(5). In particular, the gauge symmetry
breaking and the doublet-triplet splitting are discussed.  In
Sec.~\ref{sec:massmatrix} we construct the fermion mass matrices from
the invariance under the gauge symmetry as well as the discrete
symmetries.  The masses and mixing angles of quarks and 
leptons, including neutrinos, are discussed in
Sec.~\ref{sec:massmixing}.  We compare the prediction of our model
with the experimental data.  In Sec.~\ref{sec:discussion} the impact on
the leptogenesis of our model is considered. Finally,
Sec.~\ref{sec:conclusions} contains our conclusion.

%%%%%%%%%%%%%%%%%%%%%%%%%%%%%%%%%%%%%%%%%%%%%%%%%%%%%%%
\section{\label{sec:model}%
SU(5)$_1 \times$SU(5)$_2 \times$SU(5)$_3$ model}
%%%%%%%%%%%%%%%%%%%%%%%%%%%%%%%%%%%%%%%%%%%%%%%%%%%%%%%%%%%%%%%%%%%%%
\subsection{Gauge Symmetry Breaking}
%%%%%%%%%%%%%%%%%%%%%%%%%%%%%%%%%%%%%%%%%%%%%%%%%%%%%%%%%%%%%%%%%%%%%
In this article we will investigate a supersymmetric model based on a
product group:
\begin{eqnarray}
  \label{eq:G}
  G = \mbox{SU(5)}_1 \times \mbox{SU(5)}_2 \times
  \mbox{SU(5)}_3 \,.
\end{eqnarray}
One of the advantages in this class of models based on product groups
is that the gauge symmetry breaking can be achieved without
introducing Higgs fields in the adjoint or higher-dimensional
representations.  In fact, as pointed out in
Ref.~\cite{Barbieri:1994jq}, the bifundamental fields are sufficient
for this task.  We introduce here three vector-like pairs of the
bifundamental fields, $(T_1, \ol T_1)$, $(D_1, \ol D_1)$ and $(V_2,
\ol V_2)$.  Their gauge charges are presented in Table~\ref{tab:FC}.
Although $T_1 (\ol T_1)$ and $D_1 (\ol D_1)$ have the same gauge
charges, but they carry different charges under discrete symmetries.
The breaking of the group $G$ into the standard model group
SU(3)$\times$SU(2)$\times$U(1) is achieved by the vacuum expectation
values (VEVs) of these fields:
\begin{subequations}
  \label{eq:VEV_BFH}
  \begin{eqnarray}
    \vev{T_1} &=& \vev{\ol T_1} 
    = \mbox{diag}(\, T_1\,, ~T_1 \,,~ T_1\,,~0\,,~0 \,) \,,
    \\
    \vev{D_1} &=& \vev{\ol D_1} 
    = \mbox{diag}(\,0\,,~0\,,~0\,,~D_1\,,~ D_1\,) \,,
    \\
    \vev{V_2} &=& \vev{\ol V_2} 
    = \mbox{diag}(\, T_2\,,~ T_2\,,~ T_2\,,~ D_2\,,~ D_2\,) \,,
\end{eqnarray}
\end{subequations}
where we take all the VEVs are real and positive.%
%%%%%%%%%%%%%%%%%%
\footnote{We use sometimes the same letters for the superfields and
  for their VEVs.}
%%%%%%%%%%%%%%%%%
These bifundamental fields play also an important r{\^o}le for 
fermion masses and mixing.  As we will see in
Sec.~\ref{sec:massmixing}, the observational data suggest the scale
of the VEVs is typically of the order of $0.1~M_\ast$ ($M_\ast$ is the
cut-off scale which is taken as the reduced Planck scale in this
analysis).  Below these scales, we obtain the minimal supersymmetric
standard model.

The above VEVs might be determined by a superpotential for the
bifundamental fields.  Such a superpotential should avoid the
appearance of the massless Nambu-Goldstone fields.  The standard model
group is embedded in a diagonal SU(5) group of $G$, and the gauge
coupling unification is ensured at the leading order, although we have
to include the threshold corrections around the $G$ breaking scale.
In this article, we will not discuss these issues, but we
assume the VEVs as in Eqs.~(\ref{eq:VEV_BFH}).
%
%%%%%%%%%%%%%%%%%%%%%%%%%%%%%%%%%%%%%%%%%%%%%%%%%%%%%%%
\begin{table}[tb]
  \caption{\label{tab:FC}%
    Field contents of the model.}
  $
  \begin{array}{| c | c c c | c | c |}
    \hline
    {} & \mbox{SU(5)}_1 & \mbox{SU(5)}_2 & \mbox{SU(5)}_3 & Z_7 & Z_2
    \\ \hline
    T_1      & \mathbf{5} & \mathbf{1} & \mathbf{5^\ast} & 1 & -
    \\
    \ol T_1  & \mathbf{5^\ast} & \mathbf{1} & \mathbf{5} & 6 & -
    \\
    D_1      & \mathbf{5} & \mathbf{1} & \mathbf{5^\ast} & 2 & -
    \\
    \ol D_1  & \mathbf{5^\ast} & \mathbf{1} & \mathbf{5} & 5 & +
    \\
    V_2      & \mathbf{1} & \mathbf{5} & \mathbf{5^\ast} & 0 & -
    \\
    \ol V_2  & \mathbf{1} & \mathbf{5^\ast} & \mathbf{5} & 0 & -
    \\ \hline
    H        & \mathbf{1} & \mathbf{1}      & \mathbf{5} & 2 & +
    \\
    \ol H    & \mathbf{5^\ast} & \mathbf{1} & \mathbf{1} & 4 & -
    \\ \hline
    10_1     & \mathbf{10} & \mathbf{1}  &    \mathbf{1} & 1 & +
    \\
    10_2     & \mathbf{1}  & \mathbf{10} &    \mathbf{1} & 6 & +
    \\
    10_3     & \mathbf{1}  & \mathbf{1}  &   \mathbf{10} & 6 & +
    \\
    5_1^\ast & \mathbf{1} &\mathbf{5^\ast}  & \mathbf{1} & 3 & +
    \\
    5_2^\ast & \mathbf{1} & \mathbf{1} & \mathbf{5^\ast} & 2 & +
    \\
    5_3^\ast & \mathbf{1} & \mathbf{1} & \mathbf{5^\ast} & 2 & +
    \\ 
    1_1      & \mathbf{1} & \mathbf{1} &      \mathbf{1} & 2 & +
    \\
    1_2      & \mathbf{1} & \mathbf{1} &      \mathbf{1} & 3 & +
    \\ 
    1_3      & \mathbf{1} & \mathbf{1} &      \mathbf{1} & 3 & +
    \\ \hline
  \end{array}
  $
\end{table}
%%%%%%%%%%%%%%%%%%%%%%%%%%%%%%%%%%%%%%%%%%%%%%%%%%%%%%%

%%%%%%%%%%%%%%%%%%%%%%%%%%%%%%%%%%%%%%%%%%%%%%%%%%%%%%%%%%%%%%%%%%%%%
\subsection{Doublet-Triplet Splitting}
%%%%%%%%%%%%%%%%%%%%%%%%%%%%%%%%%%%%%%%%%%%%%%%%%%%%%%%%%%%%%%%%%%%%%
One of the serious difficulties in construction of GUT models is the
doublet-triplet splitting in Higgs masses.  There exist two
weak-doublets of Higgs fields, $H_u$ and $H_d$, in the minimal
supersymmetric standard model, where $H_u$ and $H_d$ are superfields
whose VEVs give masses to up- and down-type quarks, respectively.  We
incorporate these Higgs doublets by introducing
\begin{eqnarray}
  H = ( G_u \,, H_u)^T \,,~~ \ol H = ( G_d^c \,, H_d)^T \,,
\end{eqnarray}
together with color-triplet Higgs $G_u$ and $G_d^c$.  Here $H$ and
$\ol H$ are $\mathbf{5}$ and $\mathbf{5^\ast}$-plets of SU(5)$_3$ and
SU(5)$_1$, respectively. It should be emphasized that the gauge
symmetry $G$ forbids an invariant mass term for $H$ and $\ol H$,
however, the terms $\ol H T_1 H$ and $\ol H D_1 H$ are allowed in the
superpotential.  Although the former term is desirable to give a heavy
mass to the color-triplet Higgs, the latter should be forbidden to
leave the weak-doublet Higgs massless below the $G$ breaking scale.
How can we distinguish the weak-doublet Higgs from the color-triplet
Higgs in a natural way?

The doublet-triplet splitting can be
realized by using a discrete symmetry in the framework of the product-group
unification~\cite{Barbieri:1994jq,Barbieri:1994cx,Barr:1996kp,%
  Chou:1998pr,Witten:2001bf,Dine:2002se}. Here we shall follow the
discussion of Ref.~\cite{Dine:2002se}. Such a symmetry can be
constructed from a usual discrete symmetry $Z_N$ and a discrete
``hypercharge'' subgroup of U(1)$_1 \subset$ SU(5)$_1$ or U(1)$_3
\subset$ SU(5)$_3$.  In the present model we choose that of U(1)$_1$
for example.  Its element (combining with the $Z_N$ transformation)
is parametrized as
\begin{eqnarray}
  g_{1} = \mbox{diag}(\, \alpha^{-1}\,,\;
  \alpha^{-1}\,,\; \alpha^{-1}\,,\; 
  \alpha^{\frac{N+3}{2}}\,,\;
  \alpha^{\frac{N+3}{2}}\, ) \,,
\end{eqnarray}
where $\alpha$ is an $N$-th root of unity.  For the later purpose we
consider only an odd integer $N$.
We assign the $Z_N$ charges for the bifundamental fields as
\begin{eqnarray}
  \begin{array}{l l}
    T_1 \rightarrow \alpha \, T_1 \,, &
    D_1 \rightarrow \alpha^{- \frac{N+3}{2}} \, D_1 \,,
    \\
    \ol T_1 \rightarrow \alpha^{-1} \, \ol T_1 \,,~~~~ &
    \ol D_1 \rightarrow \alpha^{ \frac{N+3}{2}} \, \ol D_1 \,,
  \end{array}
\end{eqnarray}
and $V_2$ and $\ol V_2$ carry zero charge.  Then, it turns out that
the VEVs shown in Eq.~(\ref{eq:VEV_BFH}) preserve a $\widetilde Z_N$ symmetry
which is the combination of $Z_N$ and $g_{1}$.

This unbroken discrete symmetry $\widetilde Z_N$ can realize the 
doublet-triplet splitting. We assign the $Z_N$ charges for
$H$ and $\ol H$:
\begin{eqnarray}
  H \rightarrow \alpha^{n_H} \, H \,,~~~
  \ol H \rightarrow \alpha^{- n_H -1} \, \ol H\,,
\end{eqnarray}
with an integer $n_H$ ($=0,\cdots,N-1$), and the transformations
under the $\widetilde Z_N$ of each components are given by
\begin{eqnarray}
  \begin{array}{l l}
    H_u \rightarrow \alpha^{n_H} \, H_u \,, &
    G_u \rightarrow \alpha^{n_H} \, G_u \,,
    \\
    H_d \rightarrow \alpha^{- n_H - \frac{N+5}{2} } \, H_d \,,~~~~ &
    G_d^c \rightarrow \alpha^{- n_H} \, G_d^c \,.
  \end{array}
\end{eqnarray}
We can see that the $\widetilde Z_N$ symmetry, which is left after the
$G$ breaking, discriminates between the weak-doublet $H_d$ and the
color-triplet $G_d^c$ as long as $N \neq 5$.  In fact, if it is the
case, the term $\ol H D_1 H$ will be absent in the superpotential, while
the following term is still allowed
\begin{eqnarray}
  \label{eq:W_CTH}
  W \supset \ol H T_1 H \,,
\end{eqnarray}
which generates a heavy mass of $T_1\sim 0.1 M_\ast$ 
to the color-triplet Higgs.

The exact $\widetilde Z_N$ symmetry leads to no mass term for the
weak-doublet Higgs in the superpotential.  The so-called $\mu$-term
which is needed for the electroweak symmetry breaking is generated
only by its breaking.  For this purpose we introduce a gauge-singlet
field $\Phi$ which carries a $Z_N$ charge $+1$ and includes the
following term (henceforth we take the cut-off scale $M_\ast$ to be
one)
\begin{eqnarray}
  \label{eq:W_effMU}
  W \supset c_\mu \, \ol H D_1 H \, \Phi^{\frac{N+5}{2}} \,,
\end{eqnarray}
where $c_\mu$ is a constant and we set $c_\mu =1$ for simplicity.
The effective $\mu$-parameter then becomes%
\footnote{Although the effects of the supersymmetry breaking
may also give an additional contribution to 
$\mu$, it can be neglected when the supersymmetry 
breaking field carries the $Z_N$ charge zero.}
\begin{eqnarray}
 \label{eq:MuParameter}
  \mu = D_1 \, \Phi^{\frac{N+5}{2}} \,.
\end{eqnarray}
For examples, when $\mu = 100$ GeV and $D_1 = 0.1$, 
the VEV of $\Phi$ is given by $\Phi = 1.4 \times 10^{-4}$ ($N=3$),
$2.7 \times 10^{-3}$ ($N=7$) and $6.3 \times 10^{-3}$ ($N=9$).

%%%%%%%%%%%%%%%%%
\subsection{Matter Fields}
%%%%%%%%%%%%%%%%%
Now we turn to discuss the matter fields in our model.  In the
simplest SU(5) model one family of quarks and leptons are grouped into
a $\mathbf{10}$-plet and a $\mathbf{5^\ast}$-plet.  Similarly, we
introduce the matter fields as $\mathbf{10}$-plets and
$\mathbf{5^\ast}$-plets of an SU(5) factor of $G$, which guarantees
the electric charge quantization.  Concretely speaking, $10_i$,
$5_i^\ast$ and $1_i$ ($i=1,2,3$) are introduced as the matter fields.
The charges under the gauge group $G$ can be also found in
Table~\ref{tab:FC}.  First, each $10_i$ is allocated to a
$\mathbf{10}$-plet of SU(5)$_i$, respectively.  Then, the gauge
charges of $H$ and $\ol H$ results in a unique choice for the
distribution of $5_i^\ast$ to cancel the gauge anomalies.  Namely, one
of $5_i^\ast$, say $5_1^\ast$, should be a $\mathbf{5^\ast}$-plet of
SU(5)$_2$ and the rest, $5_2^\ast$ and $5_3^\ast$, should be
$\mathbf{5^\ast}$-plets of SU(5)$_3$.  Finally, three gauge-singlet
fields $1_i$ (right-handed neutrinos) are introduced to generate
neutrino masses.  

This matter content leads to interesting features of the model: First,
apart from the Majorana masses for right-handed neutrinos, the gauge
symmetry $G$ allows a Yukawa coupling only for top-quark.  All other
Yukawa couplings are induced effectively by the $G$ breaking and are
suppressed by some powers of VEVs of the bifundamental fields.  This
accounts for, at the first approximation, a large top-quark mass and smaller
masses for the other fermions.

Second, our model offers naturally the so-called ``lopsided family
structure''~\cite{Sato:1997hv,Albright:1998vf,Irges:1998ax,Barr:2000ka}.
The distribution of $10_i$ under $G$ is different from that of
$5^\ast_i$ to cancel the gauge anomalies.  This may be the reason why
there is a difference in the mass hierarchies between up- and
down-type quarks.  More importantly, the same gauge charge between
$5^\ast_2$ and $5^\ast_3$ may ensure a large $\nu_\mu$-$\nu_\tau$
mixing angle confirmed in the
atmospheric neutrino oscillation.%
\footnote{The same structure of the matter and Higgs fields (but the
  different structure of the bifundamental fields) can be found in
  Ref.~\cite{Chou:1998pr}.  However, the different mass matrices are
  induced and also the implication to the neutrino properties are not
  discussed there.}

%%%%%%%%%%%%%%%%%%%%%%%%%%%%%%%%%%%%%%%%%%%%%%%%%%%%%%%%%%%%%%%%%%%%%
%%%%%%%%%%%%%%%%%%%%%%%%%%%%%%%%%%%%%%%%%%%%%%%%%%%%%%%%%%%%%%%%%%%%%
\section{\label{sec:massmatrix}%
{}Fermion Mass Matrices}
%%%%%%%%%%%%%%%%%%%%%%%%%%%%%%%%%%%%%%%%%%%%%%%%%%%%%%%%%%%%%%%%%%%%%
%%%%%%%%%%%%%%%%%%%%%%%%%%%%%%%%%%%%%%%%%%%%%%%%%%%%%%%%%%%%%%%%%%%%%

In this section we present fermion mass matrices in our model.  As
discussed above, the product gauge group $G$ is considered as the
flavor symmetry. The discrete symmetry $Z_N$ for the doublet-triplet
splitting is also important for fermion masses.  Since the
bifundamental fields, $H$ and $\ol H$ carry non-trivial $Z_N$ charges,
the matter fields should also transform non-trivially under the $Z_N$.
Some Yukawa couplings are forbidden and induced effectively by its breaking
together with some power of the VEV of the field $\Phi$.  This means
that the $Z_N$ symmetry is crucial not only for the doublet-triplet
splitting but also for the hierarchies in fermion masses.

In this article we present the model based on the $Z_7$ symmetry
($N=7$).  This is because, first, the doublet-triplet splitting is
realized only if $N \neq 5$.  Second, when $N=3$, the VEV of $\Phi$
is too small to explain the charged fermion masses [see the discussion
below Eq.~(\ref{eq:MuParameter})].  Finally, the larger $N$, $\ie$, the
larger VEV of $\Phi$, is disfavored since it induces too small
neutrino masses.  In the following, therefore, we will
construct the fermion mass matrices based on $(i)$ the gauge symmetry
$G$ and $(ii)$ the $Z_7$ symmetry for the doublet-triplet splitting.
Further, we will impose $(iii)$ an additional $Z_2$ symmetry.  The
reason for its inclusion will be clear later.

%%%%%%%%%%%%%%%%%
\subsection{Mass Matrices for Charged Fermions}
%%%%%%%%%%%%%%%%%
Let us first consider the Yukawa terms for charged fermions,
and we will discuss neutrino masses later.
We define the Yukawa couplings by
\begin{eqnarray}
  W = Y_u{}_{ij}\, H  \, u_i^c u_j +
  Y_d{}_{ij} \, \ol H \, d_i^c d_j +
  Y_e{}_{ij} \, \ol H \, e_i^c e_j \,,
\end{eqnarray}
where $u_i$ and $u_i^c$ are the left- and right-handed up-type
quarks, $\etc$ Without taking into account the discrete symmetries,
the gauge charges of matter fields give the effective Yukawa
couplings:
\begin{subequations}
\label{eq:Yukawa_G}
\begin{eqnarray}
 Y_u &=& 
  \left( \begin{array}{c c c}
   D_1 & 
    \ol T_1{}^2 \, \ol D_2 \, \ol T_2 &
    \ol T_1{}^2
    \\
   \ol D_1 \, \ol T_1 \, \ol T_2{}^2 &
    D_2 &
    \ol T_2{}^2
    \\
   \ol D_1 \, \ol T_1 &
    \ol D_2 \, \ol T_2 &
    1
   \end{array} \right) \,,~~~~~~~
  \\
 Y_d &=& 
 \left( \begin{array}{c c c} 
  \ol T_1 \, T_2 &
   D_1 \, \ol D_2 &
   D_1 \, T_2 
   \\
  \ol T_1 &
   D_1 \, \ol D_2 \, \ol T_2 &
   D_1 \,
   \\
  \ol T_1 &
   D_1 \, \ol D_2 \, \ol T_2 &
   D_1 \,
  \end{array} \right) \,,
 \\
 Y_e{}^T &=& 
 \left( \begin{array}{c c c} 
  \ol D_1 \, D_2 &
   D_1 \, \ol D_2 &
   D_1 \, D_2 
   \\
  \ol D_1 &
   D_1 \, \ol D_2{}^2 &
   D_1 \,
   \\
  \ol D_1 &
   D_1 \, \ol D_2{}^2 &
   D_1 \,
  \end{array} \right) \,,
\end{eqnarray}
\end{subequations}
where the bifundamental fields should be considered as their diagonal
components which have non-zero VEVs (cf. $\vev{T_2} = \vev{\ol T_2} =
T_2$ and $\vev{D_2} = \vev{\ol D_2} = D_2$).  These results are
obtained at the leading order of the power expansion in the
bifundamental fields.  As we will study in Sec.~\ref{sec:massmixing},
all VEVs of the bifundamental fields turn out to be of order $0.1$ (in
the unit $M_\ast =1$), and hence the leading terms are sufficient for
our discussion.

It should be noted that we have omitted dimensionless coefficients in
front of each element in the above Yukawa couplings.  In principle, we
can take them as any complex numbers.  Here, however, following to the
idea of the Froggatt-Nielsen mechanism~\cite{Froggatt:1978nt}, we
assume that these unknown coefficients are of order one, and thus the
hierarchy in the Yukawa couplings are generated by the breaking of the
product gauge group $G$ (as well as the discrete symmetries, see
below).  Therefore, we should keep in mind that our analysis is of
order of magnitude wise and receives the uncertainty from our
ignorance to determine these coefficients.

As mentioned above, the $Z_7$ symmetry for the doublet-triplet
splitting gives rise an additional hierarchical structure in the
Yukawa couplings (\ref{eq:Yukawa_G}).  Due to the smallness of the
suppression factor, $\Phi \sim 10^{-3}$, we have to choose carefully
the $Z_7$ charges for matter fields as well as Higgs fields in order
to obtain realistic fermion mass matrices.  Notice that the $Z_7$
charges for bifundamental fields have already determined 
in Sec.~\ref{sec:model}.

Let us start with the charge assignment of the third family
fermions.
{}For a large top-quark mass we require the following term in the
superpotential without any suppression factor of $\Phi$
\begin{eqnarray}
 W \supset H \, 10_3 \, 10_3 \,,
\end{eqnarray}
and also we need the term, 
\begin{eqnarray}
 W \supset \ol H \, D_1 \, 10_3 \, 5_3^\ast \,,
\end{eqnarray}
to explain bottom and tau masses.
Moreover, the neutrino mass would come from the term,
$W \supset H \, H \,  5_{3}^\ast \, 5_{3}^\ast$,
which, unfortunately, induces too small neutrino mass scale
compared with the neutrino oscillation experiments.  From this reason, we
require the following term as a consequence of
the seesaw mechanism~\cite{seesaw,seesaw1}:%
\footnote{The same assignment of the $Z_7$ charges might induce 
the operator $W \supset \Phi^6 H \, H \, 5_{3}^\ast 5_{3}^\ast$,
however, our model is not such a case 
as we will show in Eq.~(\ref{eq:MNL2}).}
\begin{eqnarray}
  \label{eq:DIM5NU}
 W \supset \frac{1}{\Phi}\, H \, H \, 5_{3}^\ast 5_{3}^\ast \,.
\end{eqnarray}
With these three requirements we may choose the $Z_N$ charges as
\begin{eqnarray}
  H : 2 \,,~~\ol H : 4 \,,~~
  10_3 : 6 \,,~~ 5_3^\ast : 2 \,.
\end{eqnarray}
The $Z_7$ charges for other matter fields are found as follows: We
find from $m_\mu$ and $m_s$ that the charges for $10_2$ and $5_2^\ast$
should be same as $10_3$ and $5_3^\ast$, respectively.  Of course, we
have to avoid too small electron mass when the charges of $10_1$ and
$5_1^\ast$ are determined.  The results are
\begin{eqnarray}
  10_1 : 1 \,,~~ 10_2 : 6 \,,~~ 5_1^\ast : 3 \,, ~~5_2^\ast : 2 \,.
\end{eqnarray}
These charges are also listed in Table~\ref{tab:FC}.  
With this $Z_7$ charge assignment the Yukawa couplings take the forms:
\begin{subequations}
\label{eq:Yukawa_Z7}
\begin{eqnarray}
 Y_u &=& 
 \left( \begin{array}{c c c}
   \Phi \, D_1 & 
   \ol T_1{}^2 \, \ol D_2 \, \ol T_2 &
   \ol T_1{}^2
   \\
   \Phi \, \ol D_1 \, \ol T_1 \, \ol T_2{}^2 &
   D_2 &
   \ol T_2{}^2
   \\
   \Phi \, \ol D_1 \, \ol T_1 &
   \ol D_2 \, \ol T_2 &
   1
 \end{array} \right) \,,~~~~~~~
\\
Y_d &=& 
\left( \begin{array}{c c c} 
    \ol T_1 \, T_2 &
    \Phi^6 \, D_1 \, \ol D_2 &
    \Phi^6 \, D_1 \, T_2 
    \\
    \Phi \, \ol T_1 &
    D_1 \, \ol D_2 \, \ol T_2 &
    D_1 \,
    \\
    \Phi \, \ol T_1 &
    D_1 \, \ol D_2 \, \ol T_2 &
    D_1 \,
  \end{array} \right) \,,
\\
Y_e{}^T &=& 
\left( \begin{array}{c c c} 
    \Phi \, \ol D_1 \, D_2 &
    \Phi^6 \, D_1 \, \ol D_2 &
    \Phi^6 \, D_1 \, D_2 
    \\
    \Phi^2 \, \ol D_1 &
    D_1 \, \ol D_2{}^2 &
    D_1 \,
    \\
    \Phi^2 \, \ol D_1 &
    D_1 \, \ol D_2{}^2 &
    D_1 \,
  \end{array} \right) \,.
\end{eqnarray}
\end{subequations}
Comparing with Eqs.~(\ref{eq:Yukawa_G}) 
there appear suppression factors in some Yukawa couplings.

We find, however, the Yukawa couplings in Eqs.~(\ref{eq:Yukawa_Z7})
predict wrong mass ratios, $m_u/m_t$, $m_c/m_t$ and $m_d/m_b$, which
conflict with the observation by orders of magnitude.  The present
model calls for an additional discrete symmetry $Z_2$ to correct these
mass relations.  We assign non-trivial charges only for the
bifundamental and Higgs fields.  Since all matter fields carry even
parity of $Z_2$, this $Z_2$ symmetry should not be considered as a
usual horizontal flavor symmetry.  Notice that this $Z_2$ symmetry does not
affect the mass term for the color-triplet Higgs in
Eq.~(\ref{eq:W_CTH}) and the effective $\mu$-term in
Eq.~(\ref{eq:W_effMU}).
From the $Z_2$ symmetry the parity-odd components in the Yukawa 
couplings receive a suppression factor of $(D_1 \ol D_1) = \Tr (D_1 \ol D_1)
= 2 D_1^2$.  In the following analysis we neglect this factor of two 
which is beyond our approximation.

Finally, the effective Yukawa couplings in the model are given 
in the forms
\begin{subequations}
\label{eq:Yukawa_Z2}
\begin{eqnarray}
\!\!\!\!Y_u \!&\!=\!&\! 
 \left( \!\!\begin{array}{c c c}
   \Phi \, D_1 \, ( D_1 \ol D_1) & 
   \ol T_1{}^2 \, \ol D_2 \, \ol T_2 &
   \ol T_1{}^2
   \\
   \Phi \, \ol D_1 \, \ol T_1 \, \ol T_2{}^2  ( D_1 \ol D_1)&
   D_2 ( D_1 \ol D_1) &
   \ol T_2{}^2
   \\
   \Phi \, \ol D_1 \, \ol T_1  ( D_1 \ol D_1) &
   \ol D_2 \, \ol T_2 &
   1
 \end{array} \!\!\right) \,,~~~~~~~
\\
\!\!\!\!Y_d \!&\!=\!&\! 
\left( \!\!\begin{array}{c c c} 
    \ol T_1 \, T_2  ( D_1 \ol D_1) &
    \Phi^6 \, D_1 \, \ol D_2 &
    \Phi^6 \, D_1 \, T_2 
    \\
    \Phi \, \ol T_1  ( D_1 \ol D_1) &
    D_1 \, \ol D_2 \, \ol T_2 &
    D_1 \,
    \\
    \Phi \, \ol T_1  ( D_1 \ol D_1) &
    D_1 \, \ol D_2 \, \ol T_2 &
    D_1 \,
  \end{array} \!\!\right) \,,
\\
\!\!\!\!Y_e{}^T \!\!\!&\!=\!&\! 
\left( \!\!\begin{array}{c c c} 
    \Phi \, \ol D_1 \, D_2 &
    \Phi^6 \, D_1 \, \ol D_2  ( D_1 \ol D_1) &
    \Phi^6 \, D_1 \, D_2  ( D_1 \ol D_1) 
    \\
    \Phi^2 \, \ol D_1  ( D_1 \ol D_1) &
    D_1 \, \ol D_2{}^2 &
    D_1 \,
    \\
    \Phi^2 \, \ol D_1  ( D_1 \ol D_1) &
    D_1 \, \ol D_2{}^2 &
    D_1 \,
  \end{array} \!\!\right) \,.\nonumber\\
\end{eqnarray}
\end{subequations}
By putting the non-zero VEVs,
the matrices for charged fermions are obtained as follows:
\begin{subequations}
  \label{eq:MCHA}
\begin{eqnarray}
 M_u &=& H_u 
  \left( \begin{array}{c c c}
   \Phi D_1{}^3 & 
    T_1{}^2 D_2 T_2 &
    T_1{}^2
    \\
   \Phi D_1{}^3 T_1 T_2{}^2 &
    D_1{}^2 D_2 &
    T_2{}^2
    \\
   \Phi D_1{}^3 T_1 &
    D_2 T_2 &
    1
   \end{array} \right) ,~~~~~~~
  \\
 M_d &=& H_d 
 \left( \begin{array}{c c c} 
  D_1{}^2 T_1 T_2 &
   \Phi^6 D_1{}^3 D_2 &
   \Phi^6 D_1{}^3 T_2 
   \\
  \Phi T_1 &
   D_1 D_2 T_2 &
   D_1 \,
   \\
  \Phi T_1 &
   D_1 D_2 T_2 &
   D_1 \,
  \end{array} \right) ,~~~~~~~~~
 \\
 M_e{}^T &=& H_d 
 \left( \begin{array}{c c c} 
  \Phi D_1 D_2 &
   \Phi^6 D_1 D_2 &
   \Phi^6 D_1 T_2 
   \\
  \Phi^2 D_1{}^3 &
   D_1 D_2{}^2 &
   D_1 \,
   \\
  \Phi^2 D_1{}^3 &
   D_1 D_2{}^2 &
   D_1 \,
  \end{array} \right) \,.
\end{eqnarray}
\end{subequations}
It is rather important to mention that the second and third rows in
$M_d$ and $M_e{}^T$ have the same structure, since there is no
symmetry which distinguishes between $5_2^\ast$ and $5_3^\ast$. In
this analysis, however, we assume that unknown coefficients in front
of $5_2^\ast$ and $5_3^\ast$ are different. Note that some
coefficients are related of each other due to SU(5) gauge groups. For
example, the 3-3 components in $M_d$ and $M_e$ are identical.

%%%%%%%%%%%%%%%%%
\subsection{Neutrino Mass Matrices}
%%%%%%%%%%%%%%%%%
Now, we turn to discuss mass matrices for neutrinos.
We define the Dirac and the Majorana mass matrices by
\begin{eqnarray}
  W = \nu_i^c \, (M_D)_{ij} \, \nu_j +
  \frac 1 2 \, \nu_i^c \, (M_N)_{ij} \, \nu_j^c \,.
\end{eqnarray}
Since right-handed neutrinos $\nu_i^c=1_i$ are totally gauge singlets,
the gauge group $G$ tells nothing about the structure of their Majorana
masses.  On the other hand, they can carry non-trivial charges
under the discrete $Z_7$ symmetry for the doublet-triplet splitting,
and the origin of the Majorana masses and their structure may
be explained by the $Z_7$ breaking.
Moreover, this fact leads to the relation between 
the $\mu$-parameter for the Higgs mass and the neutrino masses in our model.
As we will show in the next section, this relation fits very well
to the observations.

We assign here the $Z_7$ charges for the right-handed neutrinos in the 
following way:
\begin{eqnarray}
  \label{eq:Z7_RHN}
  1_1 : 2 \,,~~ 1_2 : 3 \,,~~ 1_3 : 3 \,.
\end{eqnarray}
Then, the Majorana mass matrix becomes:
\begin{eqnarray}
  \label{eq:MNR}
 M_N &=& \Phi
 \left( \begin{array}{c c c} 
  \Phi^2 & \Phi & \Phi
   \\
  \Phi & 1 & 1
   \\
  \Phi & 1 & 1
  \end{array} \right)\,.
\end{eqnarray}
The $Z_7$ breaking generates the Majorana masses $M_i$ $(i=1,2,3)$ as
\begin{eqnarray}
  \label{eq:MNR2}
 M_1 \sim \Phi^3 \,,~~ M_{2,3} \sim \Phi \,.
\end{eqnarray}
This shows that right-handed neutrinos acquire superheavy masses
due to $\Phi = {\cal O}(10^{-3})$, hence they offer a natural
setup for the seesaw mechanism~\cite{seesaw}.

The $Z_7$ charges for right-handed neutrinos in Eq.~(\ref{eq:Z7_RHN})
are also important to construct the Dirac mass matrix $M_D$.
These charges, together with the gauge symmetry $G$ and the discrete symmetry
$Z_2$, generates $M_D$ in the form
\begin{eqnarray}
  \label{eq:MNuD}
 M_D &=& H_u
 \left( \begin{array}{c c c} 
  D_1{}^2 D_2 & \Phi & \Phi
   \\
  \Phi^6 D_1{}^2 D_2 & 1 & 1
   \\
  \Phi^6 D_1{}^2 D_2 & 1 & 1
  \end{array} \right) \,.
\end{eqnarray}
Here the suppression factor $D_1{}^2$ appears in the first row due to
the $Z_2$ symmetry.  Similar to the mass matrices for down-type quarks
and charged leptons in Eqs.~(\ref{eq:MCHA}); the second and the third
rows in $M_D$ take the same structure because $5_2^\ast$ and $5_3^\ast$
carry the same charges under all the symmetries.  These facts lead to
the large flavor mixing between the second and third families in the
left-handed leptons and right-handed down-quarks. Furthermore, we
should note that the 2-1 and 3-1 elements in $M_D$ are extremely
suppressed and essentially zeros.  These missing elements give us the
desired hierarchy in the neutrino mixing angles.  We will give these
particulars in the next section.

To summarize this section, we have constructed the mass matrices for
all quarks and leptons including neutrinos based on the product group
$G$, the $Z_7$ symmetry for the doublet-triplet splitting, and the
additional $Z_2$ symmetry.  The structure of the mass matrices
arise from the breaking of these symmetries under our assumption
that all the unknown coupling constants be of order one.
In the following section, we will compare the obtained results
with the observation.

%%%%%%%%%%%%%%%%%%%%%%%%%%%%%%%%%%%%%%%%%%%%%%%%%%%%%%%%%%%%%%%%%%%%%
%%%%%%%%%%%%%%%%%%%%%%%%%%%%%%%%%%%%%%%%%%%%%%%%%%%%%%%%%%%%%%%%%%%%%
\section{\label{sec:massmixing}%
{}Fermion Masses and Mixing}
%%%%%%%%%%%%%%%%%%%%%%%%%%%%%%%%%%%%%%%%%%%%%%%%%%%%%%%%%%%%%%%%%%%%%
%%%%%%%%%%%%%%%%%%%%%%%%%%%%%%%%%%%%%%%%%%%%%%%%%%%%%%%%%%%%%%%%%%%%%
We are now at the point to evaluate masses and mixing angles of quarks and
leptons in our model.  The fermion mass matrices are
shown in Eqs.~(\ref{eq:MCHA}), (\ref{eq:MNR}) and (\ref{eq:MNuD}).
Apart from unknown coefficients of order one,
they are all determined by the following six parameters:
\begin{eqnarray}
  \label{eq:FreeParameters}
  T_1 \,, ~~ D_1\,,~~T_2\,,~~D_2\,,~~\Phi\,,~~\tan \beta \,,
\end{eqnarray}
where $\tan\beta=H_u/H_d$, thus we expect rather non-trivial
predictions between masses and mixing angles.

%%%%%%%%%%%%%%%%%%%%%%%%%%%%%%%%%%%%%%%%%%%%%%%%%%%%%%%%%%%%%%%%%%%%%
\subsection{Charged Fermions}
%%%%%%%%%%%%%%%%%%%%%%%%%%%%%%%%%%%%%%%%%%%%%%%%%%%%%%%%%%%%%%%%%%%%%
We first discuss masses and flavor mixing of charged 
fermions. From Eqs.~(\ref{eq:MCHA}) the third family 
masses are found to be
\begin{eqnarray}
  \label{eq:MtMbMtau}
  m_t \sim H_u \,,~~ m_b \simeq m_\tau \sim H_d \, D_1 \,,
\end{eqnarray}
and we also find the following mass hierarchies:
\begin{subequations}
  \label{eq:MassRatios}
  \begin{eqnarray}
    \eqn{}\frac{m_c}{m_t} \sim D_1{}^2 \, D_2 \,,~~
    \frac{m_s}{m_b} \sim D_2 \, T_2 \,,~~
    \frac{m_\mu}{m_\tau} \sim D_2{}^2 \,,
    \\
    \eqn{}\frac{m_u}{m_t} \sim \Phi \, D_1{}^3 \,,~~
    \frac{m_d}{m_b} \sim D_1 \, T_1 \, T_2 \,,~~
    \frac{m_e}{m_\tau} \sim \Phi \, D_2 \,.~~~~~~
  \end{eqnarray}
\end{subequations}
Note that masses in the first family can be estimated in a rather
solid way since the off-diagonal elements in the first column of $M_u$ and
in the first row of $M_d$ and $M_e{}^T$ are strongly suppressed.  On the
other hand, the elements of the CKM matrix are found as
\begin{eqnarray}
 \label{eq:VCKM}
  \abs{V_{us}} \sim \frac{ \Phi \, T_1 }{ D_1 \, D_2 \, T_2 } \,,~~
  \abs{V_{cb}} \sim D_2 \, T_2 \,,~~
  \abs{V_{ub}} \sim \frac{ \Phi \, T_1 }{ D_1 } \,.
\end{eqnarray}
Here $\abs{V_{us}}$ and $\abs{V_{ub}}$ come almost from the mixing in
down-type quarks.  

It should be noted that the above relations 
between masses and mixing angles are valid at the unification scale.
Of course, we should also keep in mind that these relations
are predicted with uncertainty coming from unknown coefficients 
in the mass matrices.

As shown in Eq.~(\ref{eq:MtMbMtau}), the bottom-tau unification $m_b =
m_\tau$ in the simplest SU(5) model holds within a good accuracy even
though the mass matrices $M_d$ and $M_e{}^T$ in Eqs.~(\ref{eq:MCHA})
are different.  In addition, the unwanted SU(5) mass relations, $m_\mu
= m_s$ and $m_e = m_d$, are avoided in the model, although quarks and
leptons are unified into irreducible SU(5) representations.  This is
because of the gauge charges of matter fields as well as the structure
of our vacuum (\ref{eq:VEV_BFH}). For instance, the
Georgi-Jarlskog relation~\cite{Georgi:1979df}, $m_\mu \simeq 3 m_s$,
can be realized by taking $D_2 \simeq 3 T_2$.
Moreover, we have a factorization of the mixing angles for quarks,
\begin{eqnarray}
  \label{eq:FAC_VCKM}
\abs{V_{us}} \, \abs{V_{cb}} \sim
\abs{V_{ub}} ~\,,
\end{eqnarray}
since the quark mixing originates mostly in the structure of
the mass matrix for down-type quarks.

All the VEVs of bifundamental fields can be determined only by
the mass ratios in Eqs.~(\ref{eq:MassRatios}).  For example, without
using $m_u$ and $m_e$, the four mass ratios lead to
\begin{subequations}
  \label{eq:VEVs}
  \begin{eqnarray}
   \label{eq:VEVT1}
    T_1 \eqn{\sim} \left( \frac{m_c}{m_t} \right)^{- \frac 1 2}
    \left( \frac{m_d}{m_s} \right) 
    \left( \frac{m_\mu}{m_\tau} \right)^{\frac 3 4} 
    = 0.12\,,
    \\
   \label{eq:VEVD1}
    D_1 \eqn{\sim} \left( \frac{m_c}{m_t} \right)^{\frac 1 2}
    \left( \frac{m_\mu}{m_\tau} \right)^{- \frac 1 4} 
    = 0.098\,,
    \\
   \label{eq:VEVT2}
    T_2 \eqn{\sim} \left( \frac{m_s}{m_b} \right)
    \left( \frac{m_\mu}{m_\tau} \right)^{-\frac 1 2} 
    = 0.11\,,
    \\
   \label{eq:VEVD2}
    D_2 \eqn{\sim} \left( \frac{m_\mu}{m_\tau} \right)^{\frac 12} 
    = 0.24\,.
\end{eqnarray}
\end{subequations}
Here we have used fermion masses at the unification scale
evaluated in Ref.~\cite{Fusaoka:1998vc}. From this naive estimation
we can say that all these VEVs are typically of order $0.1$.

The rest two mass ratios give us the VEV of $\Phi$ and
a non-trivial mass relation between up-type quarks 
and charged leptons:
\begin{eqnarray}
 \label{eq:MassRelation}
 \frac{m_c^6}{m_u^4 m_t^2}
  \sim
  \frac{m_\mu^5}{m_e^4 m_\tau} ~\,.
\end{eqnarray}
Clearly this relation receives a large correction
from unknown coefficients in a very complicated way.
Additionally, there are substantial errors in $m_u$ and $m_t$
at the unification scale.
However, we may present 
the charm mass (with the highest power in the relation)
avoiding such uncertainties,
$m_c \sim 1.1$ GeV,
which is consistent with the observation within a factor $\sim 3$.
The VEV of $\Phi$ can be estimated by using $m_u$ 
\begin{eqnarray}
  \label{eq:PHI_mup}
 \Phi \sim
  \left( \frac{m_u}{m_t} \right)
  \left( \frac{m_c}{m_t} \right)^{- \frac 3 2}
  \left( \frac{m_\mu}{m_\tau} \right)^{\frac 3 4} 
  = 8.5 \times 10^{-3}\,,
\end{eqnarray}
or by using $m_e$
\begin{eqnarray}
  \label{eq:PHI_mel}
 \Phi \sim \left( \frac{m_e}{m_\tau}\right)
  \left( \frac{m_\mu}{m_\tau} \right)^{- \frac 1 2} 
  = 1.1 \times 10^{-3}\,.
\end{eqnarray}
We see that these rough estimates give $\Phi$ with a large uncertainty.

We have determined all the VEVs from the fermion mass ratios,
and hence the CKM matrix elements can be obtained as predictions.
Without the uncertainty of $\Phi$, we can evaluate
$\abs{V_{cb}}$
\begin{eqnarray}
  \abs{V_{cb}}
  \sim \frac{m_s}{m_b} = 2.7 \times 10^{-2} \,,
\end{eqnarray}
which reproduces the experimental data~\cite{Fusaoka:1998vc} with an
error smaller than a factor of two.  However, $\abs{V_{us}}$
is obtained 
\begin{eqnarray}
  \abs{V_{us}}
  \sim
  \left\{
    \begin{array}{l}
      \displaystyle
      \frac{ m_u m_t{}^{\frac 32} }{ m_c{}^{\frac 52}}
      \frac{m_\mu{}^{\frac 74}}{m_\tau{}^{\frac 74}}
      \frac{ m_d m_b }{ m_s{}^2 } = 0.40
      \\[3ex]
      \displaystyle
      \frac{m_t}{m_c} \frac{ m_e m_\mu{}^{\frac 12}}{ m_\tau{}^{\frac 32}}
      \frac{ m_d m_b }{ m_s{}^2 } = 0.054
    \end{array}
  \right. \,,
\end{eqnarray}
with a large uncertainty due to the range of $\Phi$ in
Eqs.~(\ref{eq:PHI_mup}) and (\ref{eq:PHI_mel}).  The observed value
lies indeed in this range.  The factorization relation
(\ref{eq:FAC_VCKM}) lead to $\abs{V_{ub}}$ in the range
\begin{eqnarray}
  \abs{V_{ub}} \sim (0.0014\mbox{--}0.011) \,.
\end{eqnarray}

Using the VEV of $\Phi$
we can evaluate the effective $\mu$-parameter through
Eq.~(\ref{eq:MuParameter})
\begin{eqnarray}
  \mu \sim
  \left\{
    \begin{array}{l}
      \displaystyle
      \left( \frac{m_u}{m_t} \right)^6
      \left( \frac{m_c}{m_t} \right)^{- \frac{17}{2}}
      \left( \frac{m_\mu}{m_\tau} \right)^{\frac{17}{4}}
      = 89~\mbox{TeV}\\[3ex]
      \displaystyle
      \left( \frac{m_c}{m_t} \right)^{\frac 1 2}
      \left( \frac{m_e}{m_\tau} \right)^{6}
      \left( \frac{m_\mu}{m_\tau} \right)^{- \frac{13}{4}}
      = 0.54~\mbox{GeV}
    \end{array}
  \right. \,.
\end{eqnarray}
Although the sixth power of $\Phi$ in the expression 
in Eq.~(\ref{eq:MuParameter}) enhances the uncertainly of $\Phi$,
the $\mu$-parameter of the weak scale indeed lies in this range.
This means that the breaking of the $Z_7$ symmetry
may potentially generate the desired value for $\mu$ and the charged fermion 
mass hierarchies at the same time.

To see this point more clearly, we take the effective $\mu$-term as an
input parameter, $\eg$, $\mu=100$ GeV. Then, avoiding a large
uncertainty $\Phi$ becomes from Eq.~(\ref{eq:VEVD1}):
\begin{eqnarray}
  \label{eq:VEVPHI}
 \Phi \sim \left( \frac{\mu}{D_1} \right)^{1/6}
  =  2.7 \times 10^{-3}\,.
\end{eqnarray}
In this case the mass ratios for up-quark and electron
are estimated as
\begin{eqnarray}
 &&\frac{m_u}{m_t} \sim 2.6 \times 10^{-6} \,,
 \\
 &&\frac{m_e}{m_\tau} \sim 6.6 \times 10^{-4} \,, 
\end{eqnarray}
which agree with the observed values
\begin{eqnarray}
 &&\frac{m_u}{m_t}\,\Big|_{\rm obs.} = 8.1 \times 10^{-6}\,,
 \\
 &&\frac{m_e}{m_\tau}\,\Big|_{\rm obs.} = 2.8 \times 10^{-4}\,,
\end{eqnarray}
within about a factor of three.  Notice that other mass ratios
in Eqs.~(\ref{eq:MassRatios}) are independent on $\Phi$.
We are also able to estimate the CKM elements
\begin{eqnarray}
 \abs{V_{us}} \sim 0.13\,,~~
 \abs{V_{cb}} \sim 0.027  \,,~~
 \abs{V_{ub}} \sim 0.0034 \,.
\end{eqnarray}
It is clear that we obtain rather good predictions of the quark mixing
angles, which are consistent with the current
observation~\cite{Fusaoka:1998vc} within a factor of two.
Therefore, the $\mu$-parameter of the weak scale 
and the hierarchies of charged fermions both suggest the $Z_7$
breaking scale given in Eq.~(\ref{eq:VEVPHI}).

%%%%%%%%%%%%%%%%%%%%%%%%%%%%%%%%%%%%%%%%%%%%%%%%%%%%%%%%%%%%%%%%%%%%%
\subsection{Neutrinos}
%%%%%%%%%%%%%%%%%%%%%%%%%%%%%%%%%%%%%%%%%%%%%%%%%%%%%%%%%%%%%%%%%%%%%
Next, we turn to discuss neutrino masses and mixing.  We begin
with the Majorana masses for the right-handed neutrinos.  For example,
we take here the VEV of $\Phi$ to be $2.7 \times 10^{-3}$ by taking
$\mu = 100$ GeV, as shown in Eq.~(\ref{eq:VEVPHI}).  It is then found
from Eq.~(\ref{eq:MNR}) that the Majorana masses are evaluated as
\begin{eqnarray}
  &&M_1 \sim \Phi^3 = 5.0 \times 10^{10}~\mbox{GeV} \,,
  \\
  &&M_{2,3} \sim \Phi = 6.7 \times 10^{15}~\mbox{GeV} \,.
\end{eqnarray}
With these superheavy masses the seesaw mechanism works naturally and
ensures the smallness of the effective Majorana masses for the light
neutrinos which are almost left-handed states~\cite{seesaw}.  The mass
matrix for these left-handed neutrinos is roughly estimated as
\begin{eqnarray}
  \label{eq:MNL2}
  M_\nu \simeq -M_D{}^T \, M_N{}^{-1} \, M_D
  \sim
  \frac{H_u^2}{\Phi}
  \left( \begin{array}{c c c}
      \rho^2 & \rho & \rho
      \\
      \rho & 1 & 1 
      \\
      \rho & 1 & 1
  \end{array}\right) \,,~~~
\end{eqnarray}
where we defined 
\begin{eqnarray}
  \rho \equiv \frac{D_1^2 \, D_2}{\Phi} \,.
\end{eqnarray}
Combining Eqs.~(\ref{eq:VEVD1}), (\ref{eq:VEVD2}) and
(\ref{eq:VEVPHI}) the parameter $\rho$ turns out to be $0.85$, $\ie$,
at most of order one.  The obtained result realizes the
requirement imposed in Eq.~(\ref{eq:DIM5NU}).

The particular form of the matrix $M_\nu$ given in Eq.~(\ref{eq:MNL2})
leads to interesting consequences in neutrino properties.%
\footnote{The neutrino mass matrices in the similar form have been
  discussed in the literature~\cite{Altarelli:2004za}.}  First, we
introduce the typical scale for neutrino masses $m_\nu$ which is
given by
\begin{eqnarray}
  \label{eq:TypicalMnu}
  m_\nu \equiv \frac{H_u^2}{\Phi} 
  \simeq 4.6 \times 10^{-3} \, \mbox{eV}\,.
\end{eqnarray}
The numerical estimations, which will be performed below, show that
in our model the effective neutrino mass matrix $M_\nu$ generates the
neutrino masses $m_i$ ($i=1,2,3$) with a small hierarchy
$m_3>m_2>m_1$.  We may then identify the mass squared differences
indicated by the atmospheric and solar neutrino oscillations with
$\delta m_{\rm atm}^2=m_3^2-m_2^2$ and $\delta m_{\rm
  sol}^2=m_2^2-m_1^2$. Furthermore, the typical
scale $m_\nu$ will be found to approximate the second mass $m_2$.
The recent analysis~\cite{Gonzalez-Garcia:2004it} tells us that
$m_2 \simeq \sqrt{\delta m_{\rm sol}^2} 
\simeq (7.2$--$9.9)\times10^{-3}$ eV, which is consistent with the result in
Eq.~(\ref{eq:TypicalMnu}) within a factor of two.  This result gives
us an important bridge between the neutrino masses and the electroweak
scale $\mu$-parameter.  In fact, we can find from
Eq.~(\ref{eq:MuParameter}) with $N=7$ that
\begin{eqnarray}
  \mu \eqn{=} \frac{ D_1 H_u{}^{12}}{m_\nu{}^6} \,.
\end{eqnarray}
Thus we can say that if the neutrino masses were smaller or lager by
an order of magnitude than the observed values, it would be impossible
to get the desired values for the $\mu$-parameter.  Therefore, the
neutrino mass scale in the solar neutrino oscillation
suggests a strong hint for the $\mu$-parameter of the electroweak
scale in our model.

The second consequence from Eq.~(\ref{eq:MNL2}) is that we have a
large $\nu_\mu$-$\nu_\tau$ mixing angle solution which is required by
the atmospheric neutrino experiments.  This is a direct consequence of
the fact that $5_2^\ast$ and $5_3^\ast$ have the same charges under
the gauge group $G$ as well as the discrete symmetries in order to
cancel the gauge anomalies and also to reproduce the mass hierarchies
in down-type quarks and charged leptons. 
Therefore, the model realizes the lopsided 
structure~\cite{Sato:1997hv,Albright:1998vf,Irges:1998ax,Barr:2000ka}.

Finally, we may expect that a large mixing angle for $\nu_e$-$\nu_\mu$
oscillation as well as a large $U_{e3}$ in the MNS matrix since $\rho
= {\cal O}(1)$.  In fact, one might worry that our model would give
the same result as in the so-called ``anarchy'' model for neutrino
masses~\cite{Hall:1999sn,Haba:2000be}, since the obtained mass matrix
in Eq.~(\ref{eq:MNL2}) with $\rho=1$ is identical to that in the
anarchy model.  We find, however, the following differences: $(i)$
The hypothesis in the anarchy model, the requirement of
basis-independence in neutrino flavor space, is broken in our model,
and it holds only between $\nu_\mu$ and $\nu_\tau$, which ensures a
large mixing in the atmospheric neutrino oscillations.  $(ii)$ Even if
$\rho = 1$, there are differences in the Dirac neutrino mass matrix.
Namely, as shown in Eq.~(\ref{eq:MNuD}), we have a structure from
both right- and left-handed neutrinos.  The
former one gives a negligible effect, but the latter one is important.
Due to the $Z_7$ charges of $5_i^\ast$ the 2-1 and 3-1 elements in
$M_D$ are almost zero.  Only with these missing elements, there
appears a non-trivial structure in $M_\nu$.  As we will see below, the
numerical estimates show the clear difference from the anarchy model,
and in particular there exists the small hierarchy in neutrino mixing
angles.

%%%%%%%%%%%%%%%%%%%%%%%%%%%%%%%%%%%%%%%%%%%%%%%%%%%%%%%%%%%%%%%%%%%%%
\subsection{\label{subsec:numerical}%
Numerical Estimate}
%%%%%%%%%%%%%%%%%%%%%%%%%%%%%%%%%%%%%%%%%%%%%%%%%%%%%%%%%%%%%%%%%%%%%

As we have shown above, our mass matrices can produce
phenomenologically acceptable masses and mixing angles of quarks and
leptons including neutrinos with some uncertainty from unknown
coefficients of order one.  Here we will make a more quantitative
estimate by performing the numerical calculations.

First, we include the effects of the RGE (renormalization group
equation) evolution.  For simplicity, we take the boundary at high
energy as $\Phi$ where we define all the Yukawa couplings.  Then, we
solve the one-loop RGEs down to the scale $M_Z$ and compare with the
observational data.  The threshold corrections from supersymmetric
particles are included by the universal scale $M_{\rm SUSY} = 1$ TeV.

We treat unknown coefficients of order one as follows: Our mass
matrices contain $20$ independent coefficients which are in general
complex numbers.  Note that some are related of each other due to
SU(5) groups.  These coefficients are created randomly in three
different ways (denoted by the set A, B and C).  In all sets the
phase of each coefficient is created randomly such that it is
distributed uniformly in the linear scale within the range $[0,2\pi]$.
The absolute value of each coefficient is created also randomly such that
it is distributed uniformly in the logarithmical scale within the range
$[1/1.5,1.5]$ (set A), $[1/2,2]$ (set B) and $[1/3, 3]$ (set C).
%
%%%%%%%%%%%%%%%%%%%%%%%%%%%%%%%%%%%%%%%%%%%%%%%%%%%%%%%
\begin{table}[tb]
  \caption{\label{tab:FIT}%
    VEVs of bifundamental fields and $\tan\beta$ obtained by 
minimizing $\chi^2$ in Eq.~(\ref{eq:chi2}). We also show
$\Phi$ and $\rho$ evaluated from these VEVs.}
%  \begin{ruledtabular}
  \begin{tabular}{|c | c c c c c | c c | c |} \hline
    {} & $T_1$ & $D_1$ & $T_2$ & $D_2$ & $\tan \beta$ &
    $\Phi$ & $\rho$ & $\chi^2$ \\ \hline
    Set A  & 0.188 & 0.103 & 0.0883 & 0.284 & 23.6 &
    0.00271 & 1.11 & 0.535
    \\
    Set B  & 0.175 & 0.104 & 0.0836 & 0.278 & 23.8 &
    0.00271 & 1.10 & 0.464
    \\
    Set C & 0.159 & 0.105 & 0.0768 & 0.267 & 24.0 &
    0.00270 & 1.09 & 0.404 \\ \hline
  \end{tabular} 
%  \end{ruledtabular}
\end{table}
%%%%%%%%%%%%%%%%%%%%%%%%%%%%%%%%%%%%%%%%%%%%%%%%%%%%%%%

The fermion mass matrices in our model are determined by the six
parameters as shown in Eq.~(\ref{eq:FreeParameters}).  In this
numerical study we set $\mu= 100$ GeV (at the ultraviolet boundary), for
simplicity.  There are still five free parameters -- $T_1$, $D_1$,
$T_2$, $D_2$ and $\tan \beta$ -- which have to be evaluated from the
observational data.  Here we use the masses for charged fermions $m_f$
and the CKM matrix elements $V_{\alpha \beta}$, and then the neutrino
properties are obtained as the outcome.
%
%%%%%%%%%%%%%%%%%%%%%%%%%%%%%%%%%%%%%%%%%%%%%%%%%%%%%%%
\begin{table*}
  \caption{\label{tab:MeanValues}%
    Mean values of fermion masses and mixing angles. We also 
show the observational data with $1\sigma$ error ($3\sigma$ error) 
for charged fermion masses and the CKM elements (neutrino mixing angles).}
%  \begin{ruledtabular}
   \begin{tabular}{|c|c||c|c|c|} \hline
   {} & Observation & Set A & Set B & Set C
    \\ \hline
    $m_u$ [MeV] & $2.33^{+0.42}_{-0.45}$ &
    0.859 &
    0.857 &
    0.858 
    \\
    $m_c$ [MeV] & $677^{+56}_{-61}$ &
    1.06 &
    1.05 &
    1.07 
    \\
    $m_t$ [GeV] & $174 \pm 5.5$ &
    181 &
    179 &
    174
    \\
    $m_d$ [MeV] & $4.69^{+0.60}_{-0.66}$ &
    3.71 &
    3.25 &
    2.69 
    \\
    $m_s$ [MeV] & $93.4^{+11.8}_{-13.0}$ &
    51.3 &
    51.2 &
    50.2 
    \\
    $m_b$ [GeV] & $3.00\pm0.11$ &
    2.68 &
    2.75 &
    2.91
    \\
    $m_e$ [MeV] & $0.487$ &
    0.832 &
    0.802 &
    0.760 
    \\
    $m_\mu$ [MeV] & $103$ &
    72.2 &
    74.4 &
    74.9
    \\
    $m_\tau$ [GeV] & $1.75$ &
    1.53 &
    1.57 &
    1.68 
    \\
%%%%%%%
    $\abs{V_{us}}$ & $0.220 \pm 0.0026$ &
    0.187 &
    0.187 &
    0.187 
    \\
    $\abs{V_{cb}}$ & $(4.13 \pm 0.15) \times 10^{-2}$ &
    $3.24 \times 10^{-2}$ &
    $3.26 \times 10^{-2}$ &
    $3.33 \times 10^{-2}$
    \\
    $\abs{V_{ub}}$ & $(3.67 \pm 0.47) \times 10^{-3}$ &
    $3.47 \times 10^{-3}$ &
    $3.37 \times 10^{-3}$ &
    $3.07 \times 10^{-3}$
    \\ \hline
%%%%%%%
    $\tan^2 \theta_{23}$ & 0.49--2.2 &
    1.04 &
    1.04 &
    1.05 
    \\
    $\tan^2 \theta_{12}$ & 0.29--0.64 &
    0.408 &
    0.350 &
    0.280 
    \\
    $\tan^2 \theta_{13}$ & $< 0.0571$ &
    0.118 &
    0.0954 &
    0.0608 
    \\
%%%%%%%
    $m_1$ [eV] & --- &
    $4.02 \times 10^{-4}$ &
    $3.63 \times 10^{-4}$ &
    $2.61 \times 10^{-4}$
    \\
    $m_2$ [eV] & --- &
    $3.11 \times 10^{-3}$ &
    $3.09 \times 10^{-3}$ &
    $3.08 \times 10^{-3}$
    \\
    $m_3$ [eV] & --- &
    $1.28 \times 10^{-2}$ &
    $1.39 \times 10^{-2}$ &
    $1.69 \times 10^{-2}$
    \\
%%%%%%%
    $M_1$ [GeV] & --- &
    $0.911 \times 10^{11}$ &
    $0.965 \times 10^{11}$ &
    $1.14 \times 10^{11}$ 
    \\
    $M_2$ [GeV] & --- &
    $4.36 \times 10^{15}$ &
    $4.84 \times 10^{15}$ &
    $5.34 \times 10^{15}$ 
    \\
    $M_3$ [GeV] & --- &
    $1.23 \times 10^{16}$ &
    $1.30 \times 10^{16}$ &
    $1.47 \times 10^{16}$
    \\
\hline
  \end{tabular}
%  \end{ruledtabular}
\end{table*}
%%%%%%%%%%%%%%%%%%%%%%%%%%%%%%%%%%%%%%%%%%%%%%%%%%%%%%%

We create randomly $N=10^5$ sets of the unknown coefficients, and we
calculate $m^i_f$ and $V^i_{\alpha \beta}$ for each set and take the
mean values in the logarithmical scale as follows:
\begin{eqnarray}
  \ol m_f \eqn{=} \exp \left( \frac{1}{N} \sum_{i=1}^N \ln (m^i_f) \right) \,,
  \\
  \ol V_{\alpha \beta} \eqn{=}
  \exp \left( \frac{1}{N} \sum_{i=1}^N \ln 
\abs{ V^i_{\alpha \beta}} \right) \,,
\end{eqnarray}
which should be compared with experimental data.  As we will present
later, some observables such as the CKM matrix elements are
distributed by orders of magnitude, and the mean values in the
log-scale are adequate for our analysis.

The five free parameters are fixed by minimizing the function
defined by
\begin{eqnarray}
\label{eq:chi2}
  \chi^2 \equiv\sum_{ \ol {\cal O}_I < {\cal O}_I^-}
  \left[ 
    \ln \left( \frac{\ol {\cal O}_I }{ {\cal O}_I^-} \right)
  \right]^2
  +
  \sum_{ \ol {\cal O}_I > {\cal O}_I^+}
  \left[ 
    \ln \left( \frac{\ol {\cal O}_I }{ {\cal O}_I^+} \right)
  \right]^2 \,,
\end{eqnarray}
where ${\cal O}_I$ denotes $m_f$ or 
$V_{\alpha\beta}$, and ${\cal O}_I^+$ and ${\cal O}_I^-$ 
the upper and lower limits of the observable with the $3\sigma$ error,
respectively. 
In the present analysis we use the masses at the scale $M_Z$
for charged fermions (except for top-quark) estimated in
Ref.~\cite{Fusaoka:1998vc}, and the pole mass for 
top-quark~\cite{Eidelman:wy}. Further, $\abs{V_{us}}$, $\abs{V_{cb}}$ 
and $\abs{V_{ub}}$ in Ref.~\cite{Eidelman:wy} are used for the
comparison.

The obtained values for the VEVs and $\tan\beta$ are shown in
Table~\ref{tab:FIT} for the three different sets A, B and C.  In
Table~\ref{tab:MeanValues} we also show the mean values of fermion
masses and mixing angles when $\chi^2$ takes its minimal value. We see
the results do not depend much on the choice of the sets for the
unknown coefficients.  Henceforth, we will only use the set B for our
discussions.  Note that the VEVs of bifundamental fields from
this numerical estimate are almost consistent with those shown in
Eqs.~(\ref{eq:VEVs}). We find $\tan\beta\simeq 24$.

%
%%%%%%%%%%%%%%%%%%%%%%%%%%%%%%%%%%%%%%%%%%%%%%%%%%%%%%%
\begin{figure}[t]
  \includegraphics{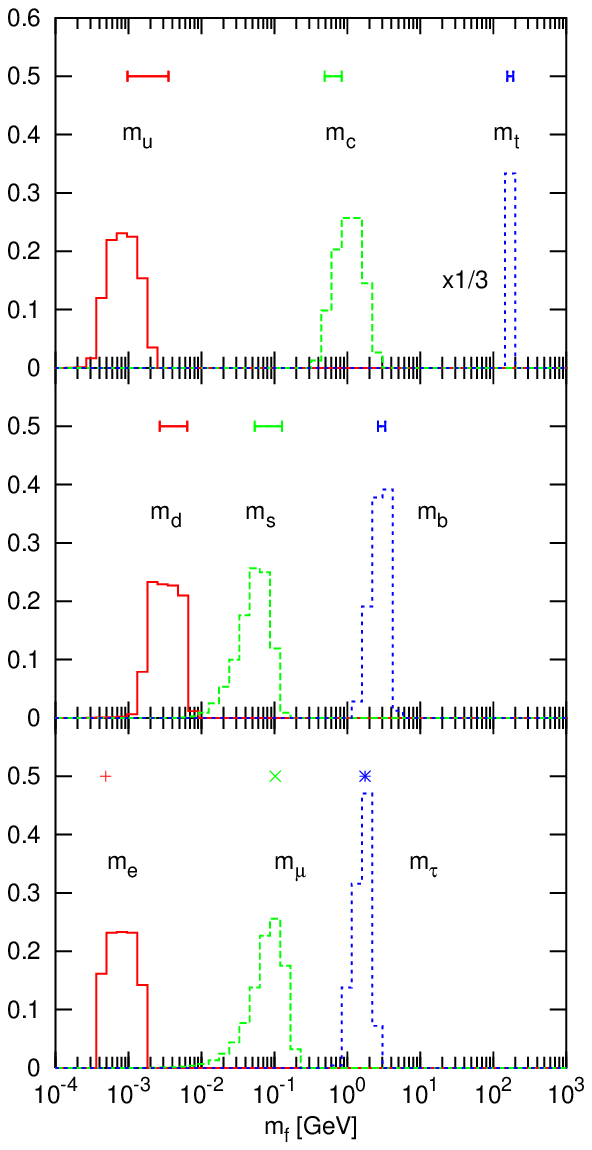}
  \includegraphics{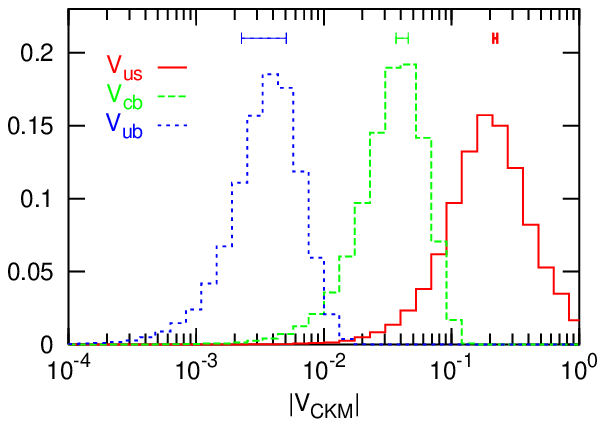}
  \caption{\label{fig:MCF_model_set2}%
    Distributions of charged fermion masses (top)
    and the CKM matrix elements (bottom). Distribution is divided
    by $3$ only for top-quark mass.  The observational data with the 
    $3\sigma$ error are also shown. 
    We use the set B for the unknown coefficients. }
\end{figure}
%%%%%%%%%%%%%%%%%%%%%%%%%%%%%%%%%%%%%%%%%%%%%%%%%%%%%%%

In Figs.~\ref{fig:MCF_model_set2} we show distributions of the charged
fermion masses and the CKM matrix elements by the $N=10^5$ trials of
creating the coefficients.  Here only the shape of the distribution
should be considered, since the hight of the histogram reflects just
the bin size of the horizontal axis.  The spread of these
distributions of masses and mixing angles originates in our ignorance
of the hidden physics to determine the unknown coefficients.

We can see that our model predicts -- of order of magnitude wise --
all masses of charged fermions as well as all quark mixing angles by
using only five parameters (apart from $\mu=100$ GeV).  One may
worry that the
model induces the wrong values for masses and mixing angles with some
possibility, since they are obtained with the distributions, $\eg$,
$\abs{V_{us}}$ distributed by two orders of magnitude.  However, the
peaks of the distributions tell that the observed values are preferred
in the model.  In this sense, the model can reproduce the charged 
fermion mass spectra and flavor mixing rather well.

%
%%%%%%%%%%%%%%%%%%%%%%%%%%%%%%%%%%%%%%%%%%%%%%%%%%%%%%%
\begin{figure}[t]
  \includegraphics{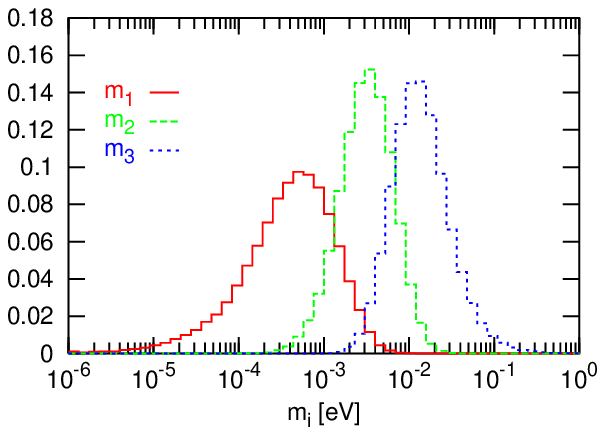}
  \includegraphics{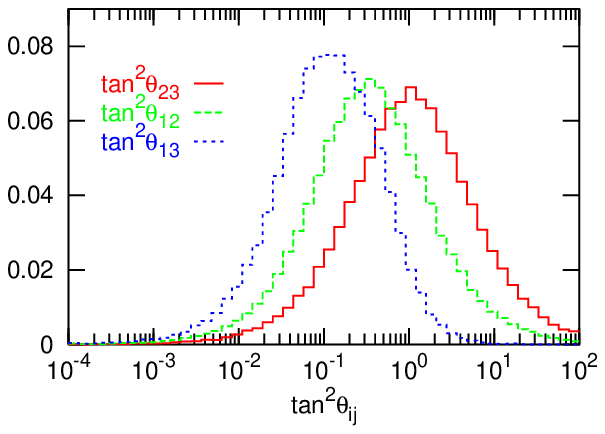}
  \caption{\label{fig:MNL_model_set2}%
    Distributions of left-handed neutrino masses $m_i$ (top) and
    neutrino mixing angles $\tan^2\theta_{ij}$ (bottom). We use the
    set B for the unknown coefficients.}
\end{figure}
%%%%%%%%%%%%%%%%%%%%%%%%%%%%%%%%%%%%%%%%%%%%%%%%%%%%%%%

Now let us discuss the neutrino masses and mixing. These quantities
are obtained as the prediction, since we have determined all free
parameters from charged fermion properties. The distributions of
left-handed neutrino masses $m_i$ are shown in
Fig.~\ref{fig:MNL_model_set2}.  It is found that there exists the
small hierarchy in neutrino masses (compared with charged fermions).
This feature comes from the same reason as in the anarchy model,
$\ie$, the multiplication of three mass matrices in the seesaw 
formula (\ref{eq:MNR2}) makes eigenvalues scattered and it is possible to
have $(m_3/m_2)\sim 10$~\cite{Hall:1999sn,Haba:2000be}.  The peaks of
the distributions are located at $m_1 \simeq 5 \times
10^{-4}~\mbox{eV}$, $m_2 \simeq 3\times 10^{-3}~\mbox{eV}$ 
and $m_3 \simeq 1 \times 10^{-2}~\mbox{eV}$.%
%%%%%%%%%%%%%%%%%%%%%%%%%%%%%%%%%%%%%%%%%%%%%%%%%%%
\footnote{We observe numerically that the peak locations of $m_1$ and
  $m_3$ (but not $m_2$) depend slightly on the choice of the unknown
  coefficients.  When we take the wider range of their norm, the peaks
  of $m_1$ and $m_3$ are shifted to the smaller and larger values,
  respectively.  The differences, however, are within a factor of
  two.}
%%%%%%%%%%%%%%%%%%%%%%%%%%%%%%%%%%%%%%%%%%%%%%%%%%%
This result shows that the most preferred value of $m_2$ corresponds
to the typical neutrino mass scale $m_\nu$ (\ref{eq:TypicalMnu}):
\begin{eqnarray}
  m_\nu = \frac{H_u^2}{\Phi} 
  = 4.6 \times 10^{-3}~\mbox{eV} \sim m_2 \,,
\end{eqnarray}
where we have used the fitted values for $\Phi$ and $\tan \beta$ in
Table~\ref{tab:FIT}.  

From the small hierarchy in the neutrino masses $m_i$
the mass squared differences in the atmospheric and solar neutrino oscillations
are given by~\cite{Gonzalez-Garcia:2004it}
\begin{eqnarray}
  \sqrt{\delta m_{\rm atm}^2} \eqn{=} (3.7\mbox{--}5.8) \times
  10^{-2}~\mbox{eV} \simeq m_3 \,,
  \\
  \sqrt{ \delta m_{\rm sol}^2 } \eqn{=} (7.2\mbox{--}9.9)\times 
  10^{-3}~\mbox{eV} \simeq m_2 \,.
\end{eqnarray}
We see that the most preferred values for $m_3$ and $m_2$ are
consistent with the observation within a factor of three or so.  With
this respect, our model is successful to explain the observed mass
spectra of neutrinos.%
\footnote{\label{fnote:drei}Although the factors of three or so are
  beyond our approximation in this analysis, we may correct them just
  by changing the overall normalization of $M_D$ or $M_N$ with a
  factor of $\sqrt{3}$ or $1/3$, which leaves other results
  unchanged.}

In Fig.~\ref{fig:MNL_model_set2} we show also the distributions of
neutrino mixing angles, $\tan^2 \theta_{ij}$.  Although the
distributions are spread by a few orders of magnitude, we may find
characteristic features in our model.  First, similar to the neutrino
masses, we find a small hierarchy in the neutrino mixing angles,
$\tan^2\theta_{23}\gtrsim\tan^2\theta_{12}\gtrsim\tan^2\theta_{13}$,
compared with the CKM angels. We should stress that this hierarchy is
obtained even when $\rho = 1.1$ [see Eq.~(\ref{eq:MNL2})].  This is
because of the missing elements in the Dirac neutrino mass matrix in
Eq.~(\ref{eq:MNuD}).

The peak of the $\tan^2\theta_{23}$ distribution suggests that it is
the maximal mixing which is confirmed in the atmospheric neutrino
oscillations.%
%%%%%%%%%%%%%%%%%%%%%%%%%%%%%%%%%%%%%%%%%%%%%%%%%%%
\footnote{It is found numerically that the peak locations of $\tan^2
  \theta_{12}$ and $\tan^2 \theta_{13}$ (but not $\tan^2 \theta_{23}$)
  depend slightly on the choice of the unknown coefficients.  The
  wider range for the norm of the coefficients makes the peaks of
  $\tan^2 \theta_{12}$ and $\tan^2 \theta_{13}$ smaller.  The
  differences, however, are within a factor of two.}
%%%%%%%%%%%%%%%%%%%%%%%%%%%%%%%%%%%%%%%%%%%%%%%%%%%
This is the direct consequence of the fact that the matter fields,
$5_2^\ast$ and $5_3^\ast$, carry the same charges for the product
group $G$ as well as the discrete symmetry $Z_7$, as mentioned before.
The peak location of the $\tan^2\theta_{12}$ distribution is found to
be $\tan^2\theta_{12}=0.35$, and hence it prefers large values,
rather than the maximal angle. This is completely different from the
anarchy model where the distributions of $\tan^2\theta_{23}$ and
$\tan^2\theta_{12}$ are the same.  

\begin{figure}[t]
  \includegraphics{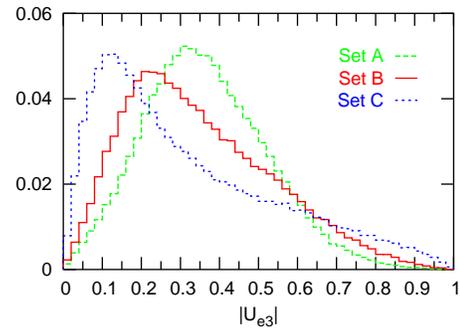}
  \caption{\label{fig:Ue3_model_comp}%
    Distributions of $\abs{U_{e3}}$ in the neutrino mixing matrix.
    We use the set A, B and C for the unknown coefficients. }
\end{figure}

Moreover, it is interesting to mention that in our model 
$\tan^2\theta_{13}<\tan^2\theta_{12}$ is preferred, and the peak location
of the $\tan^2\theta_{13}$ distribution is found to be 
$\tan^2\theta_{13}=0.1$, which is consistent with the current limit 
$\tan^2\theta_{13}<5.7\times10^{-2}$~\cite{Gonzalez-Garcia:2004it} within
a factor of two.  To be specific, we show in
Fig.~\ref{fig:Ue3_model_comp} the distributions of the $\abs{U_{e3}}$
element in the MNS matrix for the three different sets, A, B and C.
The most preferred value for $\abs{U_{e3}}$ is found to be
$0.1$--$0.3$ depending on the set.  Note that these
distributions are plotted in the linear horizontal axis, and then we
may easily see the difference between the three sets. These results
show that the forthcoming experiments on $\abs{U_{e3}}$ are
crucial for our model.
%
%%%%%%%%%%%%%%%%%%%%%%%%%%%%%%%%%%%%%%%%%%%%%%%%%%%%%%%
%%%%%%%%%%%%%%%%%%%%%%%%%%%%%%%%%%%%%%%%%%%%%%%%%%%%%%%

To summarize, our model describes well the masses of quarks and
charged leptons and also the CKM mixing angles when $\mu=100$ GeV.
Inversely, the observed mass spectra require the $\mu$-parameter
around the weak scale.  In addition, the model predicts the desired
features of neutrino properties, the correct mass scale and the small
hierarchies in both masses and mixing angles.  We should stress again
that the typical neutrino mass scale (the solar neutrino mass scale)
also indicates the weak scale $\mu$ independently on the charged
fermion sector.

%%%%%%%%%%%%%%%%%%%%%%%%%%%%%%%%%%%%%%%%%%%%%%%%%%%%%%%%%%%%%%%%%%%%%
\section{\label{sec:discussion}%
Leptogenesis}
%%%%%%%%%%%%%%%%%%%%%%%%%%%%%%%%%%%%%%%%%%%%%%%%%%%%%%%%%%%%%%%%%%%%%
%%%%%%%%%%%%%%%%%%%%%%%%%%%%%%%%%%%%%%%%%%%%%%%%%%%%%%%%%%%%%%%%%%%%%
%
%%%%%%%%%%%%%%%%%%%%%%%%%%%%%%%%%%%%%%%%%%%%%%%%%%%%%%%
\begin{figure}[t]
  \includegraphics{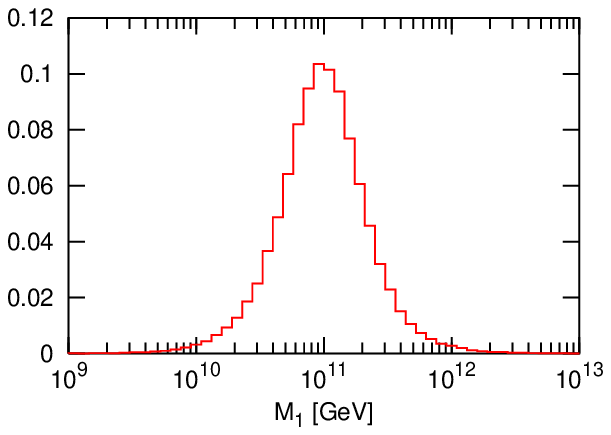}
  \includegraphics{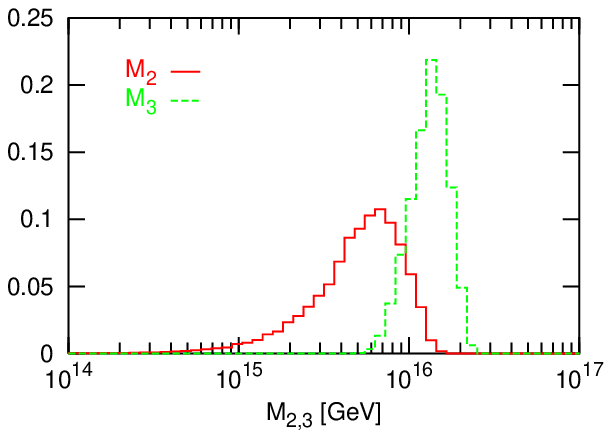}
  \caption{\label{fig:MNR1_model_set2}%
    Distributions of Majorana masses for right-handed 
    neutrinos, $M_1$ (top), $M_2$ and $M_3$ (bottom).
    We use the set B for the unknown coefficients. }
\end{figure}
%%%%%%%%%%%%%%%%%%%%%%%%%%%%%%%%%%%%%%%%%%%%%%%%%%%%%%%
%
%%%%%%%%%%%%%%%%%%%%%%%%%%%%%%%%%%%%%%%%%%%%%%%%%%%%%%%
\begin{figure}[t]
  \includegraphics{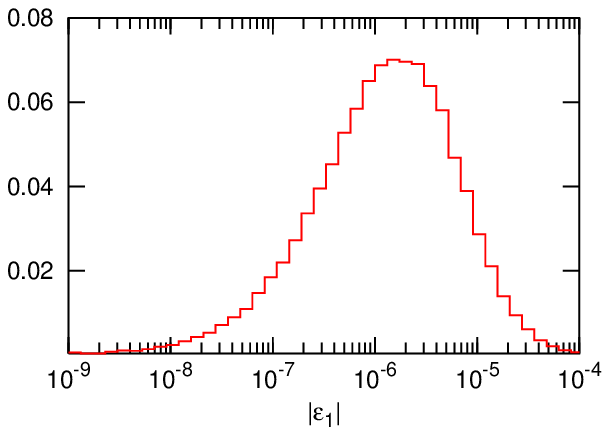}
  \includegraphics{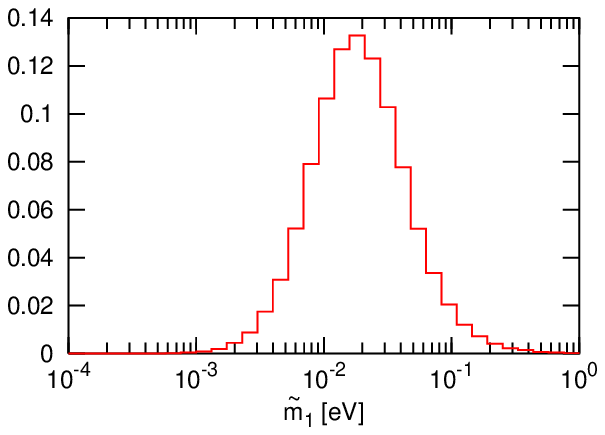}
  \caption{\label{fig:EPS1_model_set2}%
    Distributions of the CP asymmetry parameter $\abs{\epsilon_1}$
    for the lightest right-handed neutrino (top) and 
    the effective neutrino mass $\widetilde{\,m\,}_{\! 1}$ (bottom).
    We use the set B for the unknown coefficients. }
\end{figure}
%%%%%%%%%%%%%%%%%%%%%%%%%%%%%%%%%%%%%%%%%%%%%%%%%%%%%%%

As studied above, we have determined all the Yukawa couplings (or mass
matrices) of quarks and leptons including neutrinos, and hence we may
calculate various quantities in the flavor physics.  As an example,
we discuss here the implication of the model to the leptogenesis 
mechanism~\cite{Fukugita:1986hr} for the cosmic baryon asymmetry.

The non-equilibrium decays of right-handed Majorana neutrinos can
generate the lepton asymmetry, if the CP symmetry
is broken in neutrino sector, because their mass terms break the
lepton number. The CP asymmetry in the decay
of right-handed neutrino (here we consider only the 
lightest one) is parameterized by $\epsilon_1$,
which is calculated from the interference between the tree and
one-loop amplitudes for the decay processes~\cite{Luty,BuPlu,CRV}:
\begin{eqnarray}
  \epsilon_1\simeq \frac{1}{8 \pi (Y_\nu Y_\nu^\dagger)_{11}}
  \sum_{i = 2,3}
  \mbox{Im}\left[ \left\{ ( Y_\nu Y_\nu^\dagger)_{1i} \right\}^2 \right]
  \, f \left( x_i \right) \,,~~
\end{eqnarray}
where $x_i = M_i^2 / M_1^2$ and $f(x_i)$ is defined by
\begin{eqnarray}
  f(x_i) \equiv - \sqrt{x_i} \ln \left( 1 + \frac{1}{x_i} \right)
  - \frac{2 \sqrt{x_i}}{x_i - 1} \,.
\end{eqnarray}
Here and hereafter we work in the base of the Majorana mass matrix,
Eq.~(\ref{eq:MNR}), for right-handed neutrinos being diagonal.  The
distributions of the Majorana masses are
displayed in Fig.~\ref{fig:MNR1_model_set2}.  We find that typical
masses are $M_1 \simeq 10^{11}$ GeV, $M_2\simeq 7\times 10^{15}$ GeV 
and $M_3\simeq 1.3\times10^{16}$ GeV, $\ie$, $M_1\ll M_2 \lesssim M_3$.

Since we have already obtained the Dirac Yukawa couplings for
neutrinos, we are able to calculate the CP asymmetry parameter
$\epsilon_1$. We should note that since all the phases of the unknown
coefficients are distributed randomly in the range of $[0, 2\pi]$, the
sign of $\epsilon_1$ parameter can be positive or negative.  In this
analysis we neglect the sign of $\epsilon_1$ and discuss only its
absolute value.  The distribution of $\abs{\epsilon_1}$ at the scale
of $M_1$ can be found in Fig.~\ref{fig:EPS1_model_set2}:
$\abs{\epsilon_1}$ is distributed by many orders of magnitude,
however, it is seen that the most preferred
value is $\abs{\epsilon_1}\simeq2\times10^{-6}$.

This lepton asymmetry is produced when the cosmic temperature is 
$T\!\sim\! M_1$ and it is partially converted into the baryon asymmetry
through the electroweak sphaleron~\cite{Kuzmin:1985mm}.
The baryon-to-entropy ratio of the present universe is given 
by~\cite{Fukugita:1986hr,Luty,Buchmuller:2000as,Giudice:2003jh}
\begin{eqnarray}
   \frac{n_B}{s} = 1.5 \times 10^{-3} \, \kappa \, \abs{\epsilon_1} \,,
\end{eqnarray}
where $\kappa$ is the efficiency factor, which should be estimated by
solving the Boltzmann equations.  This factor depends on the lightest
Majorana mass $M_1$ and the effective neutrino mass 
$\widetilde{\,m\,}_{\! 1}$
which is defined by~\cite{Buchmuller:2000as}
\begin{eqnarray}
  \widetilde{\,m\,}_{\! 1} = \frac{ (M_D M_D^\dagger)_{11} }{M_1} \,.
\end{eqnarray}
Note that $M_D$ is the Dirac neutrino mass matrix in the basis where
right-handed neutrino mass matrix is diagonal.

Fig.~\ref{fig:EPS1_model_set2} shows also the distribution of
$\widetilde{\,m\,}_{\! 1}$ in our model, and we find
$\widetilde{\,m\,}_{\! 1}= {\cal O}(10^{-2})$ eV. According
to Ref.~\cite{Giudice:2003jh} when $M_1 = 10^{11}$ GeV and
$\widetilde{\,m\,}_{\! 1}= 2 \times 10^{-2}$ eV, $\kappa$ becomes 
about $10^{-2}$.  Therefore, the typical values 
$\abs{\epsilon_1}=2\times 10^{-6}$ and $\kappa=10^{-2}$ induce
the baryon asymmetry 
\begin{eqnarray}
  \frac{n_B}{s} \sim 3 \times 10^{-11} \,,
\end{eqnarray}
which is consistent with the recent observation
$(n_B/s)_{|{\rm obs.}}=(9.0\pm0.4)\times 10^{-11}$~\cite{Eidelman:wy}
within a factor three or so. However, the discrepancy of a factor three 
is of course beyond our estimation of order of magnitude wise,
and also there is a sizable theoretical uncertainty in
the efficiency factor~\cite{Buchmuller:2000as,Giudice:2003jh}.%
\footnote{ The correct value for the baryon asymmetry may be obtained by
  changing the overall normalization of $M_D$ (or $Y_D$) with a factor
  of $\sqrt{3}$, which also corrects the discrepancy in the light
  neutrino masses as mentioned in the footnote~\ref{fnote:drei}.}

This is a very encouraging result of the model, $\ie$, the observed
cosmic baryon asymmetry is naturally explained (with some uncertainty
from the unknown coefficients) by invoking the leptogenesis
mechanism. Unfortunately, the sign of the asymmetry is not determined
in our approach. To realize this successful scenario, the maximal
temperature of the universe should be higher than 
$M_1\simeq 10^{11}$ GeV so that the
lightest right-handed neutrinos were thermalized.  Such high
temperatures may lead to the cosmological gravitino problem in the
interesting region of the gravitino mass (of the weak
scale)~\cite{GravitinoProblem,Kawasaki:2004yh}.  To avoid this
difficulty we may go into the heavy gravitino masses of ${\cal
  O}(100)$ TeV suggested from
the anomaly mediated supersymmetry breaking mechanism~\cite{AMSB}.%
\footnote{ In Ref.~\cite{Asaka:2000ew} a possibility of having 
  high temperatures has been discussed. }
(See the discussion in Ref.~\cite{Ibe:2004tg}.)

%%%%%%%%%%%%%%%%%%%%%%%%%%%%%%%%%%%%%%
\section{\label{sec:conclusions}%
Conclusions}
%%%%%%%%%%%%%%%%%%%%%%%%%%%%%%%%%%%%%%
We have investigated in this article a supersymmetric model based on
the gauge group SU(5)$\times$SU(5)$\times$SU(5) with the discrete
symmetries $Z_7\times Z_2$.  The gauge symmetry is broken into the
standard model group only by using the bifundamental Higgs fields
without introducing the adjoint or higher-dimensional Higgs fields.
This point gives us a promising connection to the string theories. In
fact, all the fields in the model are available in the string spectrum
with the affine level one. The mass splitting between the
weak-doublets and the color-triplets of Higgs fields is ensured by the
$Z_7$ symmetry. The effective $\mu$-parameter is then generated by its 
breaking together with the gauge symmetry.

We have constructed the model for realistic masses and mixing of
quarks and leptons including neutrinos.  The hierarchical structures
in the Yukawa couplings originate in the breaking pattern of the gauge and
discrete symmetries. The fermion mass matrices are determined by only
the six parameters -- $T_1$, $D_1$, $T_2$, $D_2$, $\Phi$ and
$\tan\beta = H_u/H_d$ -- which are all the VEVs of the Higgs fields
introduced in the model.  All the masses and mixing angles are
predicted being consistent with the observation including the
uncertainty from the unknown coefficients of order unity, say, within
a factor of two or three.

There have been obtained several interesting features: First, the
bottom-tau unification, $m_b\simeq m_\tau$, is realized within a good
accuracy while avoiding the unwanted SU(5) mass relations, $m_s=m_\mu$
and $m_d=m_e$.  Second, right-handed neutrinos
carry non-trivial charges of the discrete $Z_7$ symmetry for the
doublet-triplet splitting, and its breaking generates the superheavy
Majorana masses. Thus, the model offers the natural framework for the
seesaw mechanism. Third, the $Z_7$ breaking gives us a non-trivial
bridge between three independent facts, ({\it i}) the $\mu$-parameter,
({\it ii}) the hierarchies in the charged fermion masses and the
quark-mixing angles, and ({\it iii}) the typical neutrino mass $m_\nu$
(which corresponds to the solar neutrino mass scale).  It has been
shown that both of the observational data on ({\it ii}) and ({\it iii}) 
point toward the $Z_7$ breaking scale $\Phi\simeq
2.7\times10^{-3}=6.6\times10^{15}$ GeV, which suggests interestingly
the $\mu$-parameter of order of the weak scale.  Finally, the
particular structure in the Dirac neutrino mass matrix~(\ref{eq:MNuD})
generates the small hierarchy in the neutrino mixing angles
$\tan^2\theta_{23}\gtrsim\tan^2\theta_{12} \gtrsim\tan^2\theta_{13}$.
The most preferred values of the mixing angles indicate that the
atmospheric and solar neutrino-mixing are maximal and large,
respectively.  The element $\abs{U_{e3}}$ in the MNS matrix is
typically $\abs{U_{e3}}\sim0.1$--$0.3$, which means that the present
model will be tested by the future experiments on $U_{e3}$.

%%%%%%%%%%%%%%%%%%%%%%%%%%%%%%%%%%%%%%%%%%%%%%%%%%%%%%%
We have also discussed the implication of the model to the baryon
asymmetry of the universe, $\ie$, the leptogenesis mechanism by the
decay of the lightest right-handed neutrino which is produced
thermally in the early universe.  Since the model has fixed all the
couplings and masses of neutrino (including the Majorana masses for
right-handed neutrinos), we may calculate of order of magnitude the
baryon asymmetry without any further assumptions.  Due to the random
phases of the unknown coefficients the sign of the asymmetry cannot be
determined. It has been shown that the prediction of the model
(obtained as the most preferred value) agrees with the observation
within a factor three or so, and hence the model naturally accounts
for the baryon asymmetry of the present universe. For this successful
scenario the highest temperature of the universe $T\gtrsim M_1\simeq
10^{11}$ GeV is required, and the cosmological gravitino problem
should be avoided somehow.

Before closing this article, we would like to give some comments: It
has been found from the fermion masses that the gauge symmetry
breaking scales, namely the VEVs of the bifundamental fields, are
$0.08$--$0.3\simeq(2$--$5)\times10^{17}$ GeV, which is about one
order higher than the unification scale $\simeq2\times 10^{16}$ GeV.
This might arise a trouble in the gauge coupling unification, although
it holds at the tree-level.  However, we have to take into account the
threshold corrections from superheavy particles beyond the minimal
supersymmetric standard model.  In fact, such particles are indeed
present in our model and may potentially give rise to the coupling
unification.  Moreover, we keep in mind that there might exist
non-negligible corrections from the physics beyond our model, $\eg$,
the string theories. We should also mention that the breaking of our
gauge and discrete symmetries generates the suppression for the Yukawa
couplings for the color-triplet Higgs fields and also for the
supersymmety-breaking masses for the scalar particles.  Therefore, our
model would avoid the rapid proton decay and also the supersymmetric
flavor problems.  Details of these issues will be discussed
elsewhere~\cite{TA_YT}.

\newpage
%%%%%%%%%%%%%%%%%%%%%%%%%%%%%%%%%%%%%%
\begin{acknowledgments}
  The work of TA was supported by the Tomalla foundation.
\end{acknowledgments}
%%%%%%%%%%%%%%%%%%%%%%%%%%%%%%%%%%%%%%

%\newpage
%%%%%%%%%%%%%%%%%%%%%%%%%%%%%%%%%%%%%%%%%%%%%%%%%%%%%%%

%

\begin{thebibliography}{99}
%
\bibitem{Georgi:1974sy}
H.~Georgi and S.~L.~Glashow,
%``Unity Of All Elementary Particle Forces,''
Phys.\ Rev.\ Lett.\  {\bf 32}, 438 (1974).
%%CITATION = PRLTA,32,438;%%
%
\bibitem{SUSYGUT}
S.~Dimopoulos and H.~Georgi,
%``Softly Broken Supersymmetry And SU(5),''
Nucl.\ Phys.\ B {\bf 193}, 150 (1981);
%%CITATION = NUPHA,B193,150;%%
%
N.~Sakai,
%``Naturalness In Supersymmetric 'Guts',''
Z.\ Phys.\ C {\bf 11}, 153 (1981).
%%CITATION = ZEPYA,C11,153;%%
%
\bibitem{MissingPartner}
A.~J.~Buras, J.~R.~Ellis, M.~K.~Gaillard and D.~V.~Nanopoulos,
%``Aspects Of The Grand Unification Of Strong, Weak And Electromagnetic
%Interactions,''
Nucl.\ Phys.\ B {\bf 135}, 66 (1978);
%%CITATION = NUPHA,B135,66;%%
%
H.~Georgi,
%``An Almost Realistic Gauge Hierarchy,''
Phys.\ Lett.\ B {\bf 108}, 283 (1982);
%%CITATION = PHLTA,B108,283;%%
%
A.~Masiero, D.~V.~Nanopoulos, K.~Tamvakis and T.~Yanagida,
%``Naturally Massless Higgs Doublets In Supersymmetric SU(5),''
Phys.\ Lett.\ B {\bf 115}, 380 (1982);
%%CITATION = PHLTA,B115,380;%%
%
B.~Grinstein,
%``A Supersymmetric SU(5) Gauge Theory With No Gauge Hierarchy Problem,''
Nucl.\ Phys.\ B {\bf 206}, 387 (1982).
%%CITATION = NUPHA,B206,387;%%
%

\bibitem{MissingVEV}
S.~Dimopoulos and F.~Wilczek, NSF Report No. NSF-ITP-82-07, 1981 
(unpublished);
%
M.~Srednicki,
%``Supersymmetric Grand Unified Theories And The Early Universe,''
Nucl.\ Phys.\ B {\bf 202}, 327 (1982);
%%CITATION = NUPHA,B202,327;%%
%
R.~N.~Cahn, I.~Hinchliffe and L.~J.~Hall,
%``The Hierarchy Problem In Supersymmetric Grand Unified Theories,''
Phys.\ Lett.\ B {\bf 109}, 426 (1982).
%%CITATION = PHLTA,B109,426;%%
%
\bibitem{SlidingSinglet}
E.~Witten,
%``Mass Hierarchies In Supersymmetric Theories,''
Phys.\ Lett.\ B {\bf 105}, 267 (1981);
%%CITATION = PHLTA,B105,267;%%
%
L.~E.~Ib{\'a}{\~n}ez and G.~G.~Ross,
%``SU(2)-L X U(1) Symmetry Breaking As A Radiative Effect Of Supersymmetry
%Breaking In Guts,''
Phys.\ Lett.\ B {\bf 110}, 215 (1982);
%%CITATION = PHLTA,B110,215;%%
%
D.~V.~Nanopoulos and K.~Tamvakis,
%``Susy Guts: 4 - Guts: 3,''
Phys.\ Lett.\ B {\bf 113}, 151 (1982);
%%CITATION = PHLTA,B113,151;%%
%
S.~Dimopoulos and H.~Georgi,
%``Solution Of The Gauge Hierarchy Problem,''
Phys.\ Lett.\ B {\bf 117}, 287 (1982).
%%CITATION = PHLTA,B117,287;%%
%
\bibitem{GIFT}
K.~Inoue, A.~Kakuto and H.~Takano,
%``Higgs As (Pseudo)Goldstone Particles,''
Prog.\ Theor.\ Phys.\  {\bf 75}, 664 (1986);
%%CITATION = PTPKA,75,664;%%
%
A.~A.~Anselm and A.~A.~Johansen,
%``Susy GUT With Automatic Doublet - Triplet Hierarchy,''
Phys.\ Lett.\ B {\bf 200}, 331 (1988);
%%CITATION = PHLTA,B200,331;%%
%
A.~A.~Anselm,
%``A Supersymmetric Theory Of Grand Unification With Automatic Hierarchy And
%Low-Energy Physics,''
Sov.\ Phys.\ JETP {\bf 67}, 663 (1988)
[Zh.\ Eksp.\ Teor.\ Fiz.\  {\bf 94}, 26 (1988)].
%%CITATION = SPHJA,67,663;%%
%
\bibitem{Orbifold}
Y.~Kawamura,
%``Split multiplets, coupling unification and extra dimension,''
Prog.\ Theor.\ Phys.\  {\bf 105}, 691 (2001)
[arXiv:hep-ph/0012352];
%%CITATION = HEP-PH 0012352;%%
%``Triplet-doublet splitting, proton stability and extra dimension,''
%Prog.\ Theor.\ Phys.\  
{\it ibid.} {\bf 105}, 999 (2001)
[arXiv:hep-ph/0012125];
%%CITATION = HEP-PH 0012125;%%
%
L.~J.~Hall and Y.~Nomura,
%``Gauge unification in higher dimensions,''
Phys.\ Rev.\ D {\bf 64}, 055003 (2001)
[arXiv:hep-ph/0103125].
%%CITATION = HEP-PH 0103125;%%
%
\bibitem{ATM}
Y.~Fukuda {\it et al.}  [Super-Kamiokande Collaboration],
%``Evidence for oscillation of atmospheric neutrinos,''
Phys.\ Rev.\ Lett.\  {\bf 81}, 1562 (1998)
[arXiv:hep-ex/9807003];
%%CITATION = HEP-EX 9807003;%%
%
%Y.~Fukuda {\it et al.}  [Super-Kamiokande Collaboration],
%``Measurement of the flux and zenith-angle distribution of upward
%through-going muons by Super-Kamiokande,''
%Phys.\ Rev.\ Lett.\  
{\it ibid.} {\bf 82}, 2644 (1999)
[arXiv:hep-ex/9812014];
%%CITATION = HEP-EX 9812014;%%
%
T.~Futagami {\it et al.}  [Super-Kamiokande Collaboration],
%``Observation of the east-west anisotropy of the atmospheric neutrino  flux,''
%Phys.\ Rev.\ Lett.\  
{\it ibid.}
{\bf 82}, 5194 (1999)
[arXiv:astro-ph/9901139].
%%CITATION = ASTRO-PH 9901139;%%
%
\bibitem{SOL}
S.~Fukuda {\it et al.}  [Super-Kamiokande Collaboration],
%``Solar B-8 and he p neutrino measurements from 1258 days of  Super-Kamiokande
%data,''
Phys.\ Rev.\ Lett.\  {\bf 86}, 5651 (2001)
[arXiv:hep-ex/0103032];
%%CITATION = HEP-EX 0103032;%%
%
%S.~Fukuda {\it et al.}  [Super-Kamiokande Collaboration],
%``Constraints on neutrino oscillations using 1258 days of  Super-Kamiokande
%solar neutrino data,''
%Phys.\ Rev.\ Lett.\ 
{\it ibid.}
 {\bf 86}, 5656 (2001)
[arXiv:hep-ex/0103033].
%%CITATION = HEP-EX 0103033;%%
%
\bibitem{SNO}
Q.~R.~Ahmad {\it et al.}  [SNO Collaboration],
%``Measurement of the charged current interactions produced by B-8  solar
%neutrinos at the Sudbury Neutrino Observatory,''
Phys.\ Rev.\ Lett.\  {\bf 87}, 071301 (2001)
[arXiv:nucl-ex/0106015];
%%CITATION = NUCL-EX 0106015;%%
%
%Q.~R.~Ahmad {\it et al.}  [SNO Collaboration],
%``Direct evidence for neutrino flavor transformation from neutral-current
%interactions in the Sudbury Neutrino Observatory,''
%Phys.\ Rev.\ Lett.\  
{\it ibid.} {\bf 89}, 011301 (2002)
[arXiv:nucl-ex/0204008];
%%CITATION = NUCL-EX 0204008;%%
%
%Q.~R.~Ahmad {\it et al.}  [SNO Collaboration],
%``Measurement of day and night neutrino energy spectra at SNO and constraints
%on neutrino mixing parameters,''
{\it ibid.} {\bf 89}, 011302 (2002)
[arXiv:nucl-ex/0204009];
%%CITATION = NUCL-EX 0204009;%%
%
S.~N.~Ahmed {\it et al.}  [SNO Collaboration],
%``Measurement of the total active B-8 solar neutrino flux at the Sudbury
%Neutrino Observatory with enhanced neutral current sensitivity,''
{\it ibid.}
{\bf 92}, 181301 (2004)
[arXiv:nucl-ex/0309004].
%%CITATION = NUCL-EX 0309004;%%

\bibitem{KamLAND}
K.~Eguchi {\it et al.}  [KamLAND Collaboration],
%``First results from KamLAND: Evidence for reactor anti-neutrino
%disappearance,''
Phys.\ Rev.\ Lett.\  {\bf 90}, 021802 (2003)
[arXiv:hep-ex/0212021];
%%CITATION = HEP-EX 0212021;%%
%
T.~Araki {\it et al.}  [KamLAND Collaboration],
%``Measurement of neutrino oscillation with KamLAND: Evidence of spectral
%distortion,''
arXiv:hep-ex/0406035.
%%CITATION = HEP-EX 0406035;%%

\bibitem{CHOOZ}
M.~Apollonio {\it et al.}  [CHOOZ Collaboration],
%``Limits on neutrino oscillations from the CHOOZ experiment,''
Phys.\ Lett.\ B {\bf 466}, 415 (1999)
[arXiv:hep-ex/9907037];
%%CITATION = HEP-EX 9907037;%%
%
M.~Apollonio {\it et al.},
%``Search for neutrino oscillations on a long base-line at the CHOOZ 
% nuclear power station,''
Eur.\ Phys.\ J.\ C {\bf 27}, 331 (2003)
[arXiv:hep-ex/0301017].
%%CITATION = HEP-EX 0301017;%%
%
%%%% Product GUT %%%%%%%%%%%%%%%%%%
%
\bibitem{Barbieri:1994jq}
R.~Barbieri, G.~R.~Dvali and A.~Strumia,
%``Strings versus supersymmetric GUTs: Can they be reconciled?,''
Phys.\ Lett.\ B {\bf 333}, 79 (1994)
[arXiv:hep-ph/9404278].
%%CITATION = HEP-PH 9404278;%%
%
\bibitem{Barbieri:1994cx}
R.~Barbieri, G.~R.~Dvali and A.~Strumia,
%``Fermion masses and mixings in a flavor symmetric GUT,''
Nucl.\ Phys.\ B {\bf 435}, 102 (1995)
[arXiv:hep-ph/9407239].
%%CITATION = HEP-PH 9407239;%%
%
\bibitem{Frampton:1995gu}
P.~H.~Frampton and O.~C.~W.~Kong,
%``Dicyclic Horizontal Symmetry and Supersymmetric Grand Unification,''
Phys.\ Rev.\ D {\bf 53}, 2293 (1996)
[arXiv:hep-ph/9511343].
%%CITATION = HEP-PH 9511343;%%
%
\bibitem{Mohapatra:1996fu}
R.~N.~Mohapatra,
%``$SU(5)\times SU(5)$ Unification, Seesaw Mechanism and R-Conservation,''
Phys.\ Lett.\ B {\bf 379}, 115 (1996)
[arXiv:hep-ph/9601203].
%%CITATION = HEP-PH 9601203;%%
%
\bibitem{Barr:1996kp}
S.~M.~Barr,
%``The stability of the gauge hierarchy in SU(5) x SU(5),''
Phys.\ Rev.\ D {\bf 55}, 6775 (1997)
[arXiv:hep-ph/9607359].
%%CITATION = HEP-PH 9607359;%%
%
\bibitem{Maslikov:1996gn}
A.~Maslikov, I.~Naumov and G.~Volkov,
%``The paths of unification in the GUST with the G x G gauge 
%groups of  E(8) x E(8),''
Phys.\ Lett.\ B {\bf 409}, 160 (1997)
[Mod.\ Phys.\ Lett.\ A {\bf 12}, 1909 (1997)]
[arXiv:hep-th/9612243].
%%CITATION = HEP-TH 9612243;%%
%
\bibitem{Chou:1998pr}
C.~L.~Chou,
%``Fermion mass hierarchy without flavour symmetry,''
Phys.\ Rev.\ D {\bf 58}, 093018 (1998)
[arXiv:hep-ph/9804325]; 
%%CITATION = HEP-PH 9804325;%%
%
%C.~L.~Chou,
%``Supersymmetric grand unified models without adjoint Higgs fields,''
arXiv:hep-ph/9906472.
%%CITATION = HEP-PH 9906472;%%
%
\bibitem{Witten:2001bf}
E.~Witten,
%``Deconstruction, G(2) holonomy, and doublet-triplet splitting,''
arXiv:hep-ph/0201018.
%%CITATION = HEP-PH 0201018;%%
%
\bibitem{Dine:2002se}
M.~Dine, Y.~Nir and Y.~Shadmi,
%``Product groups, discrete symmetries, and grand unification,''
Phys.\ Rev.\ D {\bf 66}, 115001 (2002)
[arXiv:hep-ph/0206268].
%%CITATION = HEP-PH 0206268;%%

\bibitem{Dienes:1996du}
K.~R.~Dienes,
%``String Theory and the Path to Unification: A Review of Recent 
%Developments,''
Phys.\ Rept.\  {\bf 287}, 447 (1997)
[arXiv:hep-th/9602045].
%%CITATION = HEP-TH 9602045;%%
%
\bibitem{Yanagida:1994vq}
T.~Yanagida,
% ``Naturally light Higgs doublets in the supersymmetric grand unified theories
%with dynamical symmetry breaking,''
Phys.\ Lett.\ B {\bf 344}, 211 (1995)
[arXiv:hep-ph/9409329].
%%CITATION = HEP-PH 9409329;%%
%
\bibitem{Froggatt:tt}
C.~D.~Froggatt, G.~Lowe and H.~B.~Nielsen,
%``The Fermion Mass Hierarchy And Gauged Chiral Flavor Quantum Numbers,''
Phys.\ Lett.\ B {\bf 311}, 163 (1993).
%%CITATION = PHLTA,B311,163;%%
%
\bibitem{Nielsen:2002cw}
H.~B.~Nielsen and Y.~Takanishi,
%``Five adjustable parameter fit of quark and lepton masses and mixings,''
Phys.\ Lett.\ B {\bf 543}, 249 (2002)
[arXiv:hep-ph/0205180], and references therein.
%%CITATION = HEP-PH 0205180;%%
%
\bibitem{Yanagida:1979gs}
T.~Yanagida,
%``Horizontal Symmetry And Mass Of The Top Quark,''
Phys.\ Rev.\ D {\bf 20}, 2986 (1979);
and see also T.~Yanagida in \cite{seesaw}.
%%CITATION = PHRVA,D20,2986;%%
%
\bibitem{Sato:1997hv}
J.~Sato and T.~Yanagida,
%``Large lepton mixing in a coset-space family unification on  
%E(7)/SU(5) x U(1)**3,''
Phys.\ Lett.\ B {\bf 430}, 127 (1998)
[arXiv:hep-ph/9710516];
%%CITATION = HEP-PH 9710516;%%
%
T.~Yanagida and J.~Sato,
%``Large lepton mixing in seesaw models: Coset-space family unification,''
Nucl.\ Phys.\ Proc.\ Suppl.\  {\bf 77}, 293 (1999)
[arXiv:hep-ph/9809307].
%%CITATION = HEP-PH 9809307;%%
%
\bibitem{Albright:1998vf}
C.~H.~Albright, K.~S.~Babu and S.~M.~Barr,
%``A minimality condition and atmospheric neutrino oscillations,''
Phys.\ Rev.\ Lett.\  {\bf 81}, 1167 (1998)
[arXiv:hep-ph/9802314];
%%CITATION = HEP-PH 9802314;%%
%
%C.~H.~Albright, K.~S.~Babu and S.~M.~Barr,
%``Implications of a minimal SO(10) Higgs structure,''
Nucl.\ Phys.\ Proc.\ Suppl.\  {\bf 77}, 308 (1999)
[arXiv:hep-ph/9805266].
%%CITATION = HEP-PH 9805266;%%
%
\bibitem{Irges:1998ax}
N.~Irges, S.~Lavignac and P.~Ramond,
%``Predictions from an anomalous U(1) model of Yukawa hierarchies,''
Phys.\ Rev.\ D {\bf 58}, 035003 (1998)
[arXiv:hep-ph/9802334].
%%CITATION = HEP-PH 9802334;%%
%
\bibitem{Barr:2000ka}
S.~M.~Barr and I.~Dorsner,
%``A general classification of three neutrino models and U(e3),''
Nucl.\ Phys.\ B {\bf 585}, 79 (2000)
[arXiv:hep-ph/0003058].
%%CITATION = HEP-PH 0003058;%%
%
\bibitem{Froggatt:1978nt}
C.~D.~Froggatt and H.~B.~Nielsen,
%``Hierarchy Of Quark Masses, Cabibbo Angles And CP Violation,''
Nucl.\ Phys.\ B {\bf 147}, 277 (1979).
%%CITATION = NUPHA,B147,277;%%
%
\bibitem{seesaw}
T.~Yanagida,
%{\it ``Horizontal Symmetry And Masses Of Neutrinos''},
%%CITATION = PTPKA,64,1103;%%
in Proceedings of the
{\it ``Workshop on the Unified Theory and the Baryon Number in the
Universe''}, Tsukuba, Japan, Feb. 13-14, 1979,
Eds. O.~Sawada and A.~Sugamoto, (KEK report KEK-79-18), p. 95;
Prog.\ Theor.\ Phys.\  {\bf 64} (1980) 1103;
P.~Ramond, Talk given at {\it ``Sanibel Symposium''}, Palm Coast, Fla.,
Feb. 25 - Mar. 2, 1979, preprint CALT-68-709;
S.~L.~Glashow, in Proceedings of the {\it ``Carg{\' e}se Summer
Insutitute on Quarks and Leptons''},  Carg{\' e}se, July 9 - 29,
1979, Eds. M.~L{\' e}vy {\it et al.}, (Plenum, 1980, New York), p. 707.
%
\bibitem{seesaw1}
R.~N.~Mohapatra and G.~Senjanovi{\'c},
%``Neutrino Mass And Spontaneous Parity Nonconservation,''
Phys.\ Rev.\ Lett.\  {\bf 44}, 912 (1980).
%%CITATION = PRLTA,44,912;%%
%
\bibitem{Georgi:1979df}
H.~Georgi and C.~Jarlskog,
%``A New Lepton - Quark Mass Relation In A Unified Theory,''
Phys.\ Lett.\ B {\bf 86}, 297 (1979).
%%CITATION = PHLTA,B86,297;%%
%
\bibitem{Fusaoka:1998vc}
H.~Fusaoka and Y.~Koide,
%``Updated estimate of running quark masses,''
Phys.\ Rev.\ D {\bf 57}, 3986 (1998)
[arXiv:hep-ph/9712201].
%%CITATION = HEP-PH 9712201;%%
%
\bibitem{Altarelli:2004za}
See, $\eg$, S.~F.~King,
%``Neutrino mass models,''
Rept.\ Prog.\ Phys.\  {\bf 67}, 107 (2004)
[arXiv:hep-ph/0310204]; 
%%CITATION = HEP-PH 0310204;%%
%
G.~Altarelli and F.~Feruglio,
%``Models of neutrino masses and mixings,''
arXiv:hep-ph/0405048.
%%CITATION = HEP-PH 0405048;%%
%
\bibitem{Gonzalez-Garcia:2004it}
For example, see a recent analysis,
M.~C.~Gonzalez-Garcia and M.~Maltoni,
%``Status of global analysis of neutrino oscillation data,''
arXiv:hep-ph/0406056.
%%CITATION = HEP-PH 0406056;%%
%
\bibitem{Hall:1999sn}
L.~J.~Hall, H.~Murayama and N.~Weiner,
%``Neutrino mass anarchy,''
Phys.\ Rev.\ Lett.\  {\bf 84}, 2572 (2000)
[arXiv:hep-ph/9911341].
%%CITATION = HEP-PH 9911341;%%
%
\bibitem{Haba:2000be}
N.~Haba and H.~Murayama,
%``Anarchy and hierarchy,''
Phys.\ Rev.\ D {\bf 63}, 053010 (2001)
[arXiv:hep-ph/0009174].
%%CITATION = HEP-PH 0009174;%%
%
\bibitem{Eidelman:wy}
S.~Eidelman {\it et al.}  [Particle Data Group Collaboration],
%``Review Of Particle Physics,''
Phys.\ Lett.\ B {\bf 592}, 1 (2004).
%%CITATION = PHLTA,B592,1;%%
%
\bibitem{Fukugita:1986hr}
M.~Fukugita and T.~Yanagida,
%``Baryogenesis Without Grand Unification,''
Phys.\ Lett.\ B {\bf 174}, 45 (1986).
%%CITATION = PHLTA,B174,45;%%
%
\bibitem{Luty}
M.~A.~Luty,
%``Baryogenesis via leptogenesis,''
Phys.\ Rev.\ D {\bf 45}, 455 (1992).
%%CITATION = PHRVA,D45,455;%%
%
\bibitem{BuPlu}
W.~Buchm{\"u}ller and M.~Pl{\"u}macher,
%``CP asymmetry in Majorana neutrino decays,''
Phys.\ Lett.\ B {\bf 431}, 354 (1998)
[arXiv:hep-ph/9710460].
%%CITATION = HEP-PH 9710460;%%
%
\bibitem{CRV}
L.~Covi, E.~Roulet and F.~Vissani,
%``CP violating decays in leptogenesis scenarios,''
Phys.\ Lett.\ B {\bf 384}, 169 (1996)
[arXiv:hep-ph/9605319].
%%CITATION = HEP-PH 9605319;%%
%
\bibitem{Kuzmin:1985mm}
V.~A.~Kuzmin, V.~A.~Rubakov and M.~E.~Shaposhnikov,
%``On The Anomalous Electroweak Baryon Number Nonconservation In The Early
%Universe,''
Phys.\ Lett.\ B {\bf 155}, 36 (1985).
%%CITATION = PHLTA,B155,36;%%
%
\bibitem{Buchmuller:2000as}
For a review, see, $\eg$,
W.~Buchm{\"u}ller and M.~Pl{\"u}macher,
%``Neutrino masses and the baryon asymmetry,''
Int.\ J.\ Mod.\ Phys.\ A {\bf 15}, 5047 (2000)
[arXiv:hep-ph/0007176].
%%CITATION = HEP-PH 0007176;%%
%
\bibitem{Giudice:2003jh}
G.~F.~Giudice, A.~Notari, M.~Raidal, A.~Riotto and A.~Strumia,
%``Towards a complete theory of thermal leptogenesis in the SM and MSSM,''
Nucl.\ Phys.\ B {\bf 685}, 89 (2004)
[arXiv:hep-ph/0310123].
%%CITATION = HEP-PH 0310123;%%
%
\bibitem{GravitinoProblem}
M.~Y.~Khlopov and A.~D.~Linde,
%``Is It Easy To Save The Gravitino?,''
Phys.\ Lett.\ B {\bf 138}, 265 (1984);
%%CITATION = PHLTA,B138,265;%%
%
J.~R.~Ellis, J.~E.~Kim and D.~V.~Nanopoulos,
%``Cosmological Gravitino Regeneration And Decay,''
Phys.\ Lett.\ B {\bf 145}, 181 (1984).
%%CITATION = PHLTA,B145,181;%%
%
\bibitem{Kawasaki:2004yh}
See, for example, a recent analysis, 
M.~Kawasaki, K.~Kohri and T.~Moroi,
%``Hadronic decay of late-decaying particles and big-bang nucleosynthesis,''
arXiv:astro-ph/0402490.
%%CITATION = ASTRO-PH 0402490;%%
%
\bibitem{AMSB}
L.~Randall and R.~Sundrum,
%``Out of this world supersymmetry breaking,''
Nucl.\ Phys.\ B {\bf 557}, 79 (1999)
[arXiv:hep-th/9810155];
%%CITATION = HEP-TH 9810155;%%
%
G.~F.~Giudice, M.~A.~Luty, H.~Murayama and R.~Rattazzi,
%``Gaugino mass without singlets,''
JHEP {\bf 9812}, 027 (1998)
[arXiv:hep-ph/9810442].
%%CITATION = HEP-PH 9810442;%%
%
\bibitem{Asaka:2000ew}
T.~Asaka and T.~Yanagida,
%``Solving the gravitino problem by axino,''
Phys.\ Lett.\ B {\bf 494}, 297 (2000)
[arXiv:hep-ph/0006211].
%%CITATION = HEP-PH 0006211;%%
%
\bibitem{Ibe:2004tg}
M.~Ibe, R.~Kitano, H.~Murayama and T.~Yanagida,
%``Viable supersymmetry and leptogenesis with anomaly mediation,''
arXiv:hep-ph/0403198.
%%CITATION = HEP-PH 0403198;%%
%
\bibitem{TA_YT}
T.~Asaka and Y.~Takanishi,
in preparation.
%
\end{thebibliography}
\end{document}